\documentclass[sigconf]{acmart}

\AtBeginDocument{%
  }

\setcopyright{none}

\acmConference[arXiv '26]{arXiv}{April, 2026}{}

\usepackage{subcaption}
\usepackage{wrapfig}
\usepackage{makecell}

\begin{document}

\title{SurgGaze: Implicit Calibration for Accurate Gaze Analysis in Operating Rooms with Wearable Eyetrackers}

\author{Jingying Wang}
\affiliation{%
  \institution{Computer Science and Engineering, University of Michigan}
  \city{Ann Arbor}
  \state{Michigan}
  \country{United States}
}
\email{wangchy@umich.edu}
\author{Rosiana Natalie}
\affiliation{%
  \institution{Michigan Institute for Data \& AI in Society, University of Michigan}
  \city{Ann Arbor}
  \state{Michigan}
  \country{United States}
}
\email{rosianan@umich.edu}

\author{Keyuan Hu}
\affiliation{%
  \institution{Computer Science and Engineering, University of Michigan}
  \city{Ann Arbor}
  \state{Michigan}
  \country{United States}
}
\email{keyuanhu@umich.edu}

\author{Wenqian Xu}
\affiliation{%
  \institution{School of Information, University of Michigan}
  \city{Ann Arbor}
  \state{Michigan}
  \country{United States}
}
\email{wtxu@umich.edu}

\author{Brian George}
\affiliation{%
  \institution{Medical School, University of Michigan}
  \city{Ann Arbor}
  \state{Michigan}
  \country{United States}
}
\email{bcgeorge@med.umich.edu}

\author{Vitaliy Popov}
\affiliation{%
  \institution{Medical School, University of Michigan}
  \city{Ann Arbor}
  \state{Michigan}
  \country{United States}
}
\email{vipopov@umich.edu}

\author{Anhong Guo}
\affiliation{%
  \institution{Computer Science and Engineering, University of Michigan}
  \city{Ann Arbor}
  \state{Michigan}
  \country{United States}
}
\email{anhong@umich.edu}

\author{Xu Wang}
\affiliation{%
  \institution{Computer Science and Engineering, University of Michigan}
  \city{Ann Arbor}
  \state{Michigan}
  \country{United States}
}
\email{xwanghci@umich.edu}

\renewcommand{\shortauthors}{Wang et al.}

\begin{abstract}

Accurate gaze tracking is essential for understanding surgeons' visual attention and cognitive processes during laparoscopic surgery, yet wearable eye trackers produce large errors systematically correlated with ground-truth gaze locations, as demonstrated in Study~1. We introduce \textit{SurgGaze}, an implicit calibration method that corrects these errors using high-confidence surgical moments. Building on evidence that surgeons' gaze converges near the tool-tissue contact point (TTCP) during dissection, \textit{SurgGaze} uses TTCP as a surrogate for true gaze to construct training pairs. We evaluate \textit{SurgGaze} in a simulated operating room trial and an authentic operating room case study. In simulation, \textit{SurgGaze} reduced gaze estimation error by 40.6\%, significantly outperforming conventional 9-point explicit calibration. The case study showed that these moments provide reliable training data and that calibrated gaze improves interpretation of surgeons' attention beyond numeric error reduction. These findings demonstrate that structured behavioral signals can enable implicit calibration for gaze tracking in complex real-world environments.

\end{abstract}



\begin{teaserfigure}
  \includegraphics[width=\textwidth]{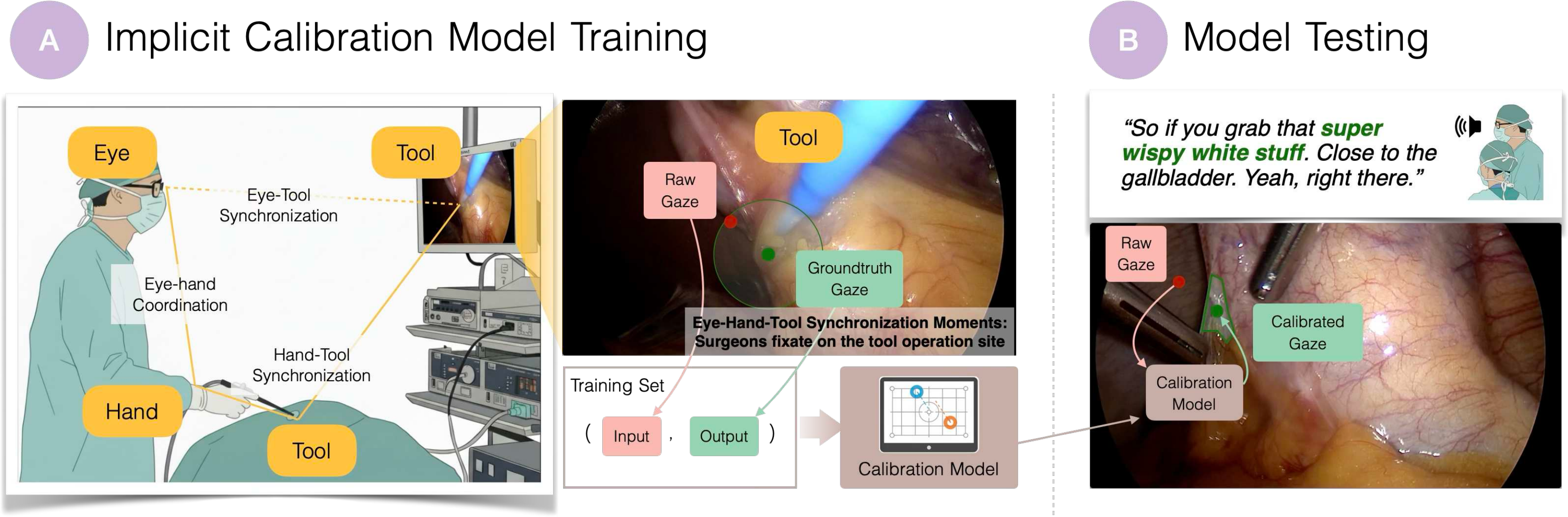}
  \caption{\textit{SurgGaze} is an implicit calibration technique for mobile eye tracking on the laparoscopic screen. The method consists of a calibration model and the training data that drives it. (A) The ground truth gaze in the training data is derived from tool-tissue contact points (TTCPs): during critical moments (e.g., dissection), surgeons fixate on where the tooltip touches the tissue. (B) Through  evaluation in authentic ORs, we show that the calibrated gaze is 8 times more likely to fall into the regions the surgeons were attending to.
  }
  \Description{}
  \label{fig:teaser}
\end{teaserfigure}

\received{20 February 2007}
\received[revised]{12 March 2009}
\received[accepted]{5 June 2009}

\maketitle
\section{Introduction}
Mobile eye tracking has become an essential tool for studying human behavior, attention, cognition, and decision-making processes~\cite{kassner2014pupil, zhang2015appearance, jiao2025hagi}, and is increasingly used in high-stakes domains where gaze provides rich behavioral and cognitive signals, such as surgery~\cite{popov2024looking, mentis2012interaction, mentis2013imaging}.
Surgery is an important HCI domain in which clinicians perform high-stakes physical actions through mediated visual interfaces~\cite{wang2024surgment, kim2023surch, chellali2016achieving, mentis2014learning}, where even small breakdowns in perception or coordination can directly affect patient safety~\cite{way2003causes,popov2024looking, bogdanova2016depth}. Most surgical training occurs on the job in the operating room (OR), where residents (trainees) and attending (faculty) surgeons coordinate to carry out procedures~\cite{TeachingHospitals, glarner2017resident, belyansky2011poor}. The OR therefore functions as both a practice and learning environment, where trainees develop both technical (e.g., instrument handling) and non-technical (e.g., communication) skills~\cite{weiser2008estimation, haynes2009surgical}. A key component of this learning is knowing where and what to look at~\cite{Vajsbaher2022The, glarner2017resident, Guzmán-García2022Correlating}. This is especially critical in minimally invasive surgery (MIS)~\cite{vajsbaher2018spatial,keehner2004spatial,de2020resident}, where visual misperception accounts for up to 97\% of errors in the most common MIS procedure, laparoscopic cholecystectomy (lap chole)~\cite{elfenbein2016confidence, kim2020mind, george2017readiness, law2004eye}. Therefore, visual attention in the OR directly impacts patient safety and the training outcomes for residents. As gaze reflects visual attention, it serves as an important performance metric in the OR~\cite{mentis2014learning, wilson2011gaze}. Both trainees and attending surgeons value capturing gaze, particularly for postoperative review and feedback on experts' attention strategies~\cite{popov2024looking}. 
 
\begin{figure}[t!]
    \centering
    \begin{subfigure}[t]{0.44\linewidth}
        \centering
        \includegraphics[width=\linewidth]{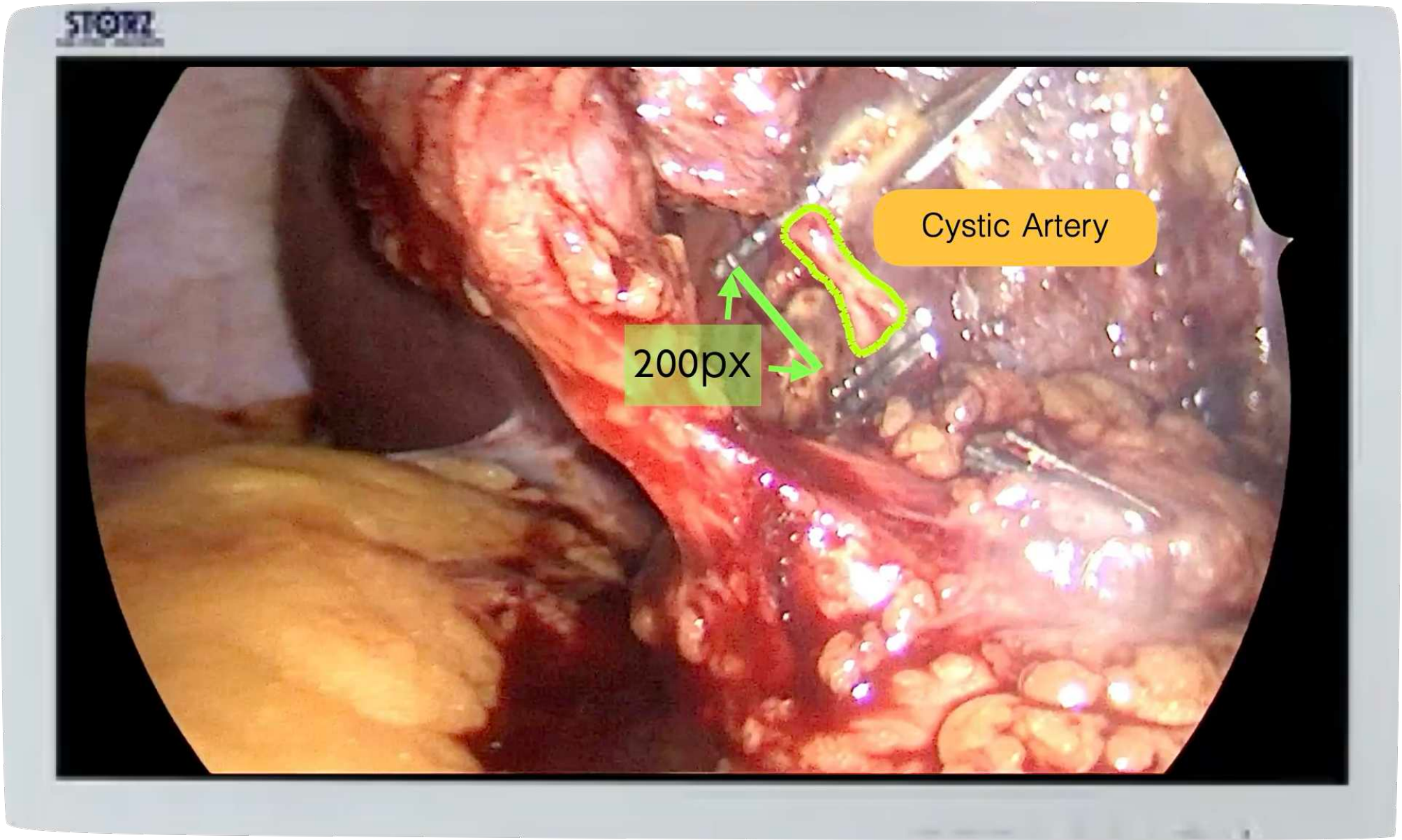}
        \caption{The cystic artery, a critical surgical landmark, often spans only about 200 pixels on the laparoscopic screen.}
        \label{fig:screen200px}
    \end{subfigure}
    \hfill
    \begin{subfigure}[t]{0.54\linewidth}
        \centering
        \includegraphics[width=\linewidth]{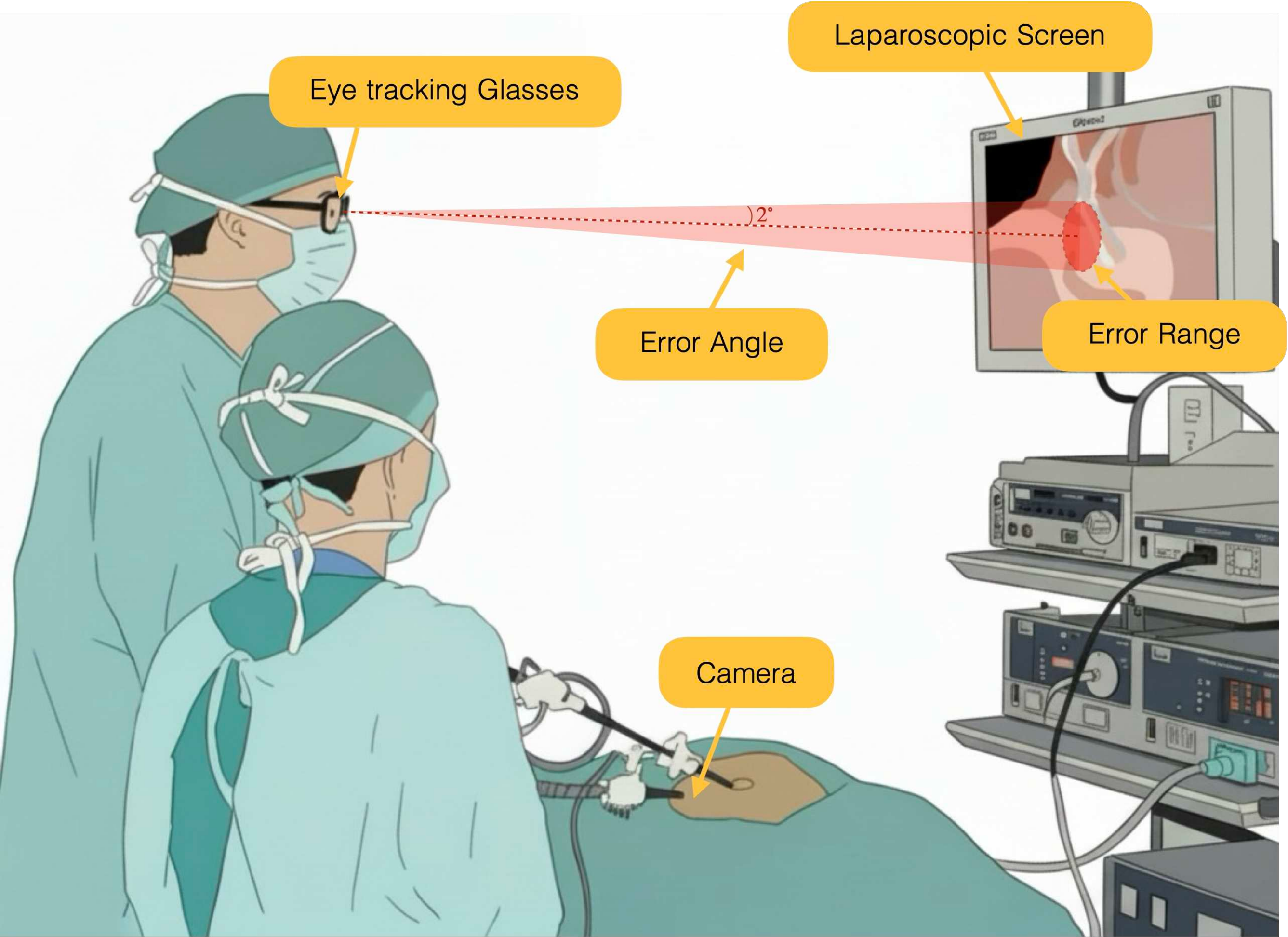}
        \caption{Eye tracking setup in laparoscopic OR. A typical state-of-the-art wearable eye tracker has $\sim$$2^\circ$ angular error for static targets, which translates to approximately 269.6 pixels on screen at 1.2\,m viewing distance.}
        \label{fig:setup}
    \end{subfigure}
    \caption{Wearable eye trackers produce errors larger than key anatomical structures, making them inadequate for accurate gaze capture in the operating rooms (OR).}
\end{figure}
In this work, we aim to \textbf{bring accurate eye tracking with wearable eye trackers to real ORs during MIS}. However, eye tracking in MIS is challenging. Surgeons operate using a video feed displayed on laparoscopic screens (Fig.~\ref{fig:teaser}), typically standing about 1.2 meters away, which makes screen-based eye trackers infeasible. Wearable eye trackers are therefore the only option~\cite{atkins2013surgeons}, but their angular error is amplified when projected onto a distant screen: at 1.2\,m, a typical $2^\circ$ angular error for static targets translates to approximately 269.6 pixels (Fig.~\ref{fig:setup}). This is problematic because many anatomical structures are small (e.g., the cystic artery spans ~200 pixels on screen; Fig.~\ref{fig:screen200px}).

We began by asking a foundational question: \textbf{(RQ1) How accurate are state-of-the-art wearable eye trackers in OR settings for MIS procedures?} We performed Study~1 ($N=14$) in a high-fidelity simulated operating room that replicates real OR layout, equipment, and workflows. This study benchmarks gaze estimation from two head-mounted eye trackers (Tobii Pro Glasses 2\footnote{\url{https://www.tobii.com/products/discontinued/tobii-pro-glasses-2}}, Pupil Labs Neon\footnote{\url{https://pupil-labs.com/products/neon}}).
The key takeaways are: (1) raw gaze output from the eye trackers has a large average error (309 px for Neon and 582 px for Tobii), exceeding the size of critical structures like the cystic artery, and (2) the raw gaze errors are strongly correlated with the position of real fixation, suggesting potential for correction via data-driven modeling.

Based on these insights, we proceeded to investigate \textbf{(RQ2) What calibration techniques could compensate for these errors?}
We propose \textit{SurgGaze}, an implicit calibration method that learns to correct systematic eye tracking errors in the OR leveraging structured surgical behaviors where ground-truth gaze can be reliably inferred. The advantage of implicit calibration is twofold. First, it avoids a separate, isolated calibration routine, which is difficult to fit into OR workflow. Second, it changes the calibration data source: instead of relying on sparse static targets collected outside the task, implicit calibration can use abundant samples collected during the task itself. For gaze analysis in surgery, this means learning calibration from dynamic laparoscopic viewing activity rather than from a brief 9-point fixation routine.

The conceptual approach of \textit{SurgGaze} is grounded in classic theories of hand-eye coordination \cite{Johansson2001EyeHand} that people almost always fixate near the point of contact between the hand and the object. In surgical contexts, prior studies report that during active dissection (i.e., the most critical phase of surgery where surgeons must precisely identify, separate, and manipulate anatomical structures) \cite{zheng2021action}, surgeons fixate on the tool-tissue contact point (TTCP), especially in two situations: (a) when the tooltip first contacts the target tissue, causing the tool and anatomical target to spatially overlap~\cite{Liu2021Developing, Li2023Using}; and (b) during energy application with a cautery tool, when surgeons must closely monitor each cut on the patient~\cite{bardakcioglu2015advanced}. These moments therefore provide a natural and reliable signal for gaze calibration.

This conceptual approach builds on prior HCI research showing a strong coupling between gaze and hand movements: users tend to look where they click or tap on screens~\cite{huang2012user,chen2013eye}. This relationship has supported both gaze-based interaction techniques~\cite{rivu2020gaze,rivu2019gazebutton,liang2024e2go} and implicit gaze calibration methods that use touchscreen behaviors such as taps and swipes~\cite{Cai2025GazeSwipe:}. 
\textit{SurgGaze} extends this line of work to a real-world surgical setting, where the visual interface is much farther from the user (surgical monitors rather than touchscreens) and the implicit calibration signals are more heterogeneous and less well-defined than discrete touch interactions. 

We evaluate \textit{SurgGaze} by separating these two claims. Study~2 first tests the data-source premise of implicit calibration: whether abundant in-task samples collected during dynamic laparoscopic target-following reduce gaze error more than sparse static samples from a conventional 9-point routine. This study uses assigned tooltip targets as ground truth, allowing us to isolate the effect of calibration sample source. It does not yet test whether TTCPs can be extracted reliably in real surgery. Study~3 then tests the OR-specific premise: whether naturally occurring tool-tissue contact points (TTCPs) can provide this kind of abundant in-task calibration signal during authentic surgical activity.

Study~2 ($N=22$) addressed \textbf{(RQ3) How effective is abundant in-task samples (implicit calibration) in simulated OR?} Participants viewed laparoscopic recordings in a simulated OR and fixated on assigned tooltip targets while wearing the Pupil Labs Neon eye tracker. We compared dense in-task calibration with conventional 9-point explicit calibration~\cite{huynh2021imon, park2021gazel, Santini2017CalibMe:, cheng2022easygaze, li2021device}. Dense in-task calibration reduced gaze error by 40.6\% (319$\rightarrow$167 px on a Full HD (FHD) screen), whereas the 9-point calibration achieved a 16.9\% reduction (319$\rightarrow$224 px). These results support the value of shifting calibration from sparse static samples to abundant in-task samples.

Study~3 then addressed \textbf{(RQ4) How effective is \textit{SurgGaze}, which uses TTCPs as abundant in-task calibration samples, in authentic OR settings?} Study~3 involved five live lap chole cases at a major U.S. teaching hospital. To extract TTCP-based calibration samples, a medical expert spent approximately 15 minutes identifying and labeling 150 tool-tissue contact frames per case. The test set was derived independently from surgeons' in-procedure verbal references, annotated as regions of interest (ROIs). Results show that \textit{SurgGaze} substantially improved gaze alignment: the proportion of gaze samples falling within the ROI increased from 11.34\% (uncalibrated) to 85.75\% (calibrated), an approximately eightfold improvement.

This work provides empirical evidence on deploying mobile eye tracking in a high-stakes environment: surgery, extending prior work on implicit gaze calibration leveraging touchscreen interactions. It also validates a broader conceptual approach: implicit calibration based on heterogeneous and structured behavioral signals can effectively improve gaze calibration in the wild, where people act on the physical world through a mediated visual interface. 
This opens up new avenues for applying implicit calibration in other real-world tasks and domains, such as teleoperation, telestration, and construction involving camera-assisted equipment. 

\section{Related Work}
\subsection{Eye tracking in the Operating Room}
Eye tracking has been widely studied in medical domains. In the OR, it is primarily used to assess surgeon expertise, guide training, and evaluate performance~\cite{khan2012analysis, he2020eye, 10.1145/968363.968370}. Gaze patterns distinguish trainees from experts. For example, experts exhibit longer fixations on task-relevant regions, less tool-following behavior, and more anticipatory gaze shifts~\cite{he2020eye}. These insights have informed training systems such as gaze overlays, video replays, and multi-view displays, which help novices develop expert-like strategies~\cite{kottayil2016investigation, Liu2021Developing, Chen2019Looks}. Gaze metrics also capture surgical skill and workload~\cite{hermens2013eye}, and can support post-operative debriefs to improve visual alignment between trainees and attendings~\cite{popov2024looking}.

While existing research demonstrates the value of gaze data for training and assessment in the OR, there remains a limited understanding of how eye tracking performs in real OR settings for MIS~\cite{atkins2013surgeons}. Most prior work falls into three categories. First, gaze tracking in controlled laboratory environments, such as simulation-box tasks for laparoscopic skill practice involving cube manipulation, does not reflect real OR layouts, workflows, or time pressure~\cite{Li2023Using, Wu2019Eye-Tracking, Le2023570, Marín-Conesa2021The}. Second, some studies rely on post-surgery video review using screen-based trackers, which do not capture gaze during live procedures~\cite{khan2012analysis, zheng2021action, tien2012measuring}. Third, real-time studies often analyze gaze using coarse metrics such as blink rate, pupil size~\cite{Li2023Using}, fixation frequency, and saccade length~\cite{Oh2024Quantitative}, without examining fine-grained object-level attention. Our work addresses this gap by investigating the performance of eye tracking devices in real OR settings and how to improve the accuracy of wearable eye trackers in the OR.

\subsection{Gaze Calibration Techniques: Explicit vs. Implicit}
Commercial eye trackers still exhibit non-negligible residual error~\cite{Holmqvist2019Peer, Barsingerhorn2018Development} when deployed in settings requiring high levels of precision, making calibration foundational. Explicit calibration protocols (e.g., 5- or 9-point targets) require users to fixate on predefined points~\cite{Garde2021Low-Cost, Wen2020Accurate} or follow moving targets~\cite{Barsingerhorn2018Development}. While effective in controlled settings, explicit calibration is impractical in the OR~\cite{atkins2013surgeons}: surgeons cannot spare time, displays cannot be altered, and frequent repositioning would demand repeated recalibration. Beyond workflow disruption, explicit calibration also limits the calibration data source. A 9-point routine provides sparse samples from a static, isolated fixation task. For downstream gaze analysis in surgery, however, the target use case is dynamic laparoscopic activity, where visual targets move, tools interact with tissue, and attention is shaped by ongoing action. Prior work on dataset shift and cross-dataset generalization shows that mismatch between training and deployment distributions can reduce model performance~\cite{QuioneroCandela2008Dataset,Torralba2011DatasetBias,Koh2021WILDS}. This motivates implicit calibration not only as a way to avoid interruption, but also as a way to collect calibration samples from within the activity where gaze will be interpreted.

Implicit calibration instead derives reference points from natural interactions such as touch events~\cite{jiang2020we, weill2016you, Cai2025GazeSwipe:, Liu2025Enhancing}, cursor clicks~\cite{huang2016building, kasprowski2016implicit, liebling2014gaze}, smooth pursuit of moving targets~\cite{pfeuffer2013pursuit}, and inferred visual saliency~\cite{yang2021vgaze}. For example, GazeSwipe~\cite{Cai2025GazeSwipe:} uses tap locations as proxies for true gaze to train correction models. However, these approaches have not been tested in surgery, where hand-eye-tool coordination introduces unique dynamics. Two questions remain open: what is the ``tapping’’ moment in laparoscopic procedures, and do these coordination patterns provide reliable calibration signals?

\subsection{Eye-Hand-Tool Coordination}
Prior work documents a strong coupling between gaze and manual input: gaze and mouse positions overlap for much of the time~\cite{liebling2014gaze}, fixations cluster around touch events on tablets~\cite{weill2016you}, and similar patterns emerge across handwriting, hovering, and browsing tasks~\cite{zhao2017eyes,weill2018correlation,bieg2010eye,rodden2008eye,milisavljevic2021similarities,milisavljevic2018eye}. During object manipulation, gaze is directed to ``obligatory’’ task landmarks such as the grasp site at the moment of hand contact~\cite{Johansson2001EyeHand}.

In surgery, a similar pattern holds. Experts fixate on surgical targets and rely on peripheral vision to monitor tools, whereas novices frequently alternate between the tool and the target~\cite{Wilson2011Perceptual,Liu2021Developing,Maeda2021Years,law2004eye,zheng2021action}. Critically, during action phases, both novices and experts tend to fixate either on the tool or on the target tissue. When the tool and target overlap, such as when the tool tip contacts tissue during dissection, the surgeon’s fixation is most likely located at that shared touch point~\cite{law2004eye,Wilson2011Perceptual}. In this paper, we exploit this convergence by modeling the tool as the cursor and the manipulated tissue as the target, using the tool tissue contact point during dissection as a natural proxy for ground truth gaze.

\section{Study 1: Characterizing Gaze Accuracy in Simulated OR Settings}
To address \textbf{(RQ1) How accurate are state-of-the-art wearable eye trackers in OR settings?}, we conducted a study benchmarking two widely used devices under conditions representative of real surgical practice in a simulated OR environment.
\subsection{Method}
This IRB-aprroved study investigates wearable eye tracker performance in a high-fidelity simulated OR at a large U.S. teaching hospital. We recruited 14 participants (8M, 6F; aged 21--35; see Table~\ref{tab:s1participants} in Appendix) selected for diversity in physiological characteristics relevant to eye tracking (e.g., eye color, eyelid anatomy, eyewear). Since the task only required following a surgical tooltip in video, no surgical expertise was needed.

\begin{figure}[t!]
    \centering
    \includegraphics[width=\linewidth]{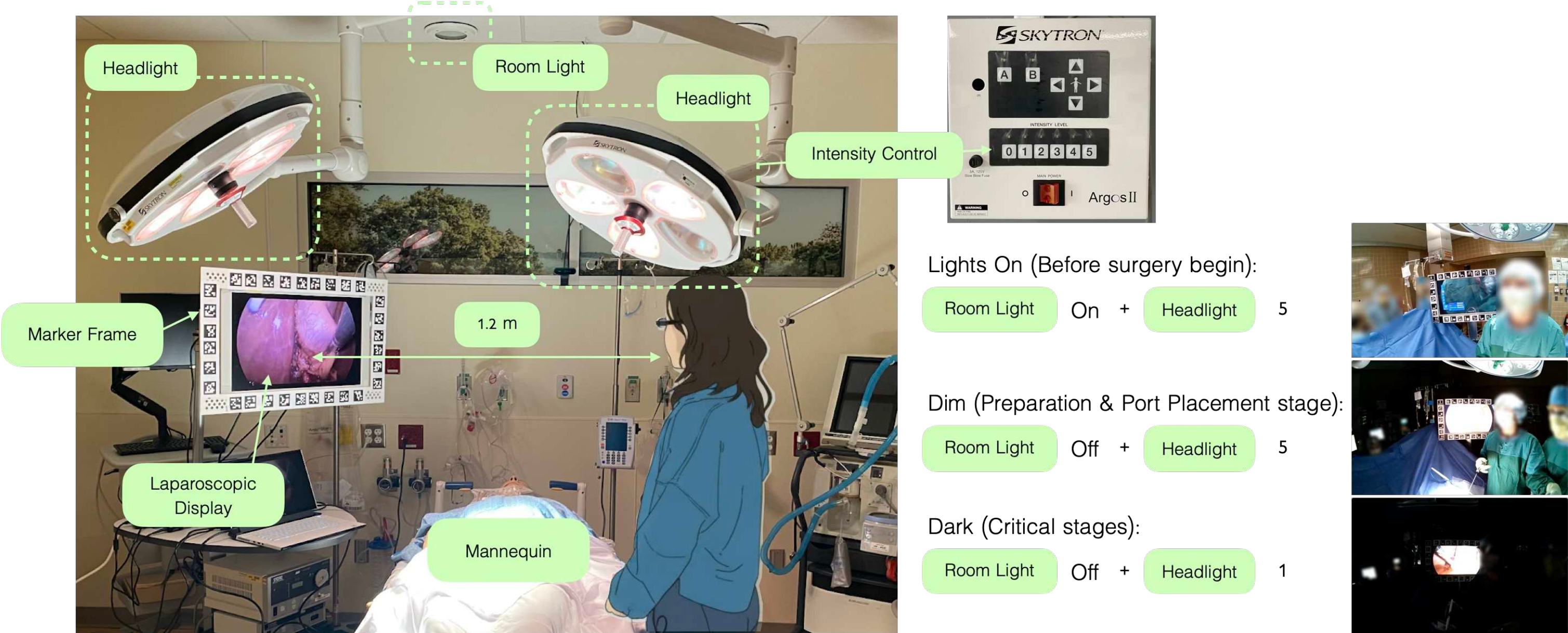}
    \caption{Experimental setup for Study 1. A 27-inch FHD laparoscopic display was positioned at 1.2\,m from the participant. Frames with markers for screen tracking were mounted on the display for frame alignment. Participants wore Tobii Pro Glasses 2 and Pupil Labs Neon eye trackers while viewing 30-second lap chole video clips. Lighting was controlled via room lights and a surgical headlight.}
    \label{fig:study1-setup}
\end{figure}

Participants stood 1.2\,m from a 27-inch FHD laparoscopic display~\cite{el2006optimum} and watched 30-second lap chole clips while wearing Tobii Pro Glasses 2 and Pupil Labs Neon (Fig.~\ref{fig:study1-setup}). Each completed 36 counterbalanced trials (2 eye trackers $\times$ 3 lighting conditions~\cite{Blignaut2013Eye-tracking} $\times$ 3 head postures~\cite{Kapp2021ARETT:}; for the full trial procedure, see Appendix Fig.~\ref{fig:s1procedure}). Sessions lasted approximately 50 minutes; participants were compensated \$25. We computed gaze error as the Euclidean distance between raw gaze and the ground truth tooltip center.

\subsection{Findings}
\subsubsection{Gaze error is too large for practical use}
\begin{figure}
    \centering
    \includegraphics[width=0.8\linewidth]{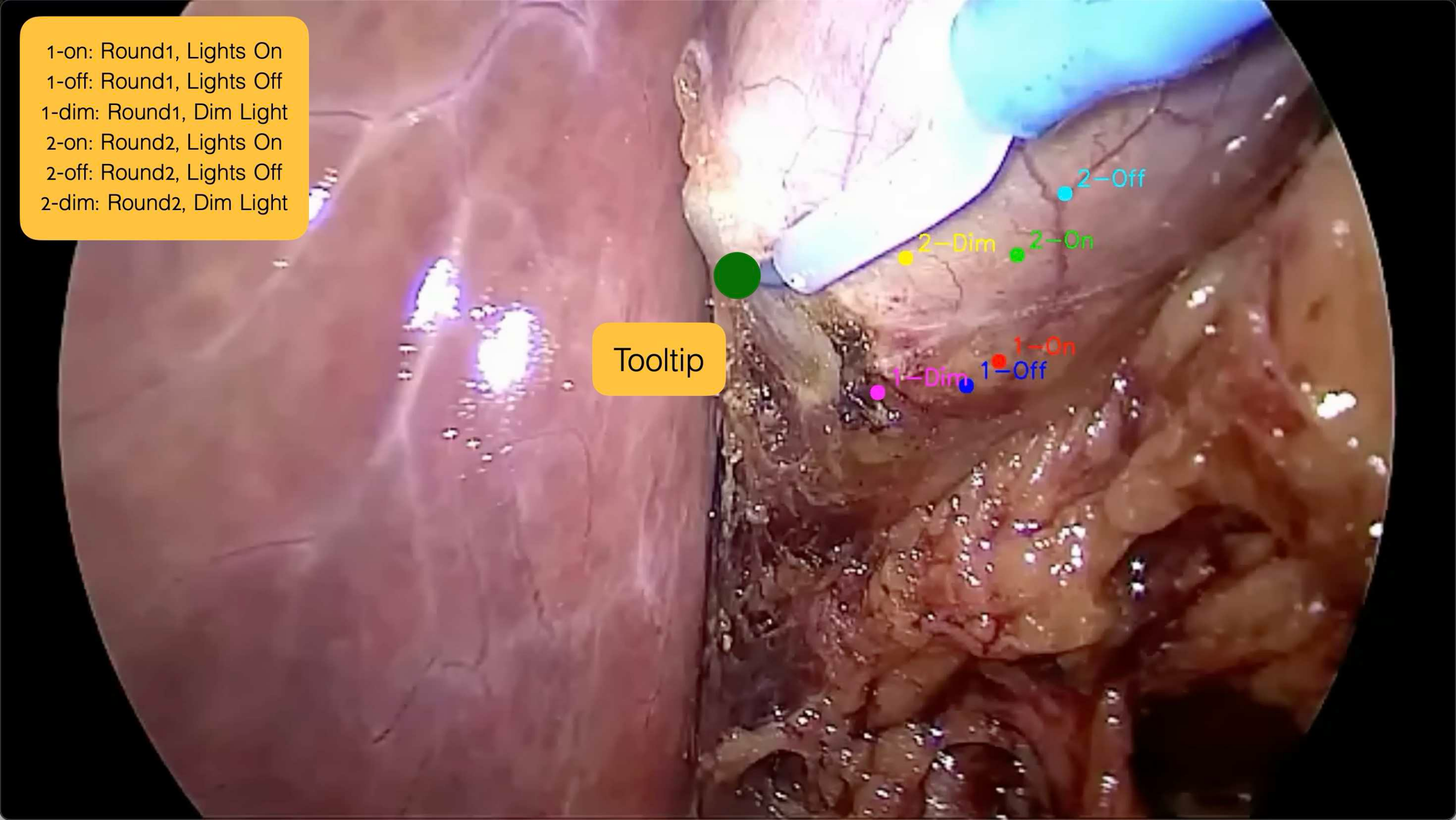}
    \caption{Example of surgical tooltip and raw gaze. We plotted the gaze samples from the Neon tracker in different rounds under multiple lighting conditions. All the raw gazes fall on the right side of the actual tool tip.}
    \label{fig:tipoutside}
\end{figure}

\begin{figure}
    \centering
    \includegraphics[width=0.5\linewidth]{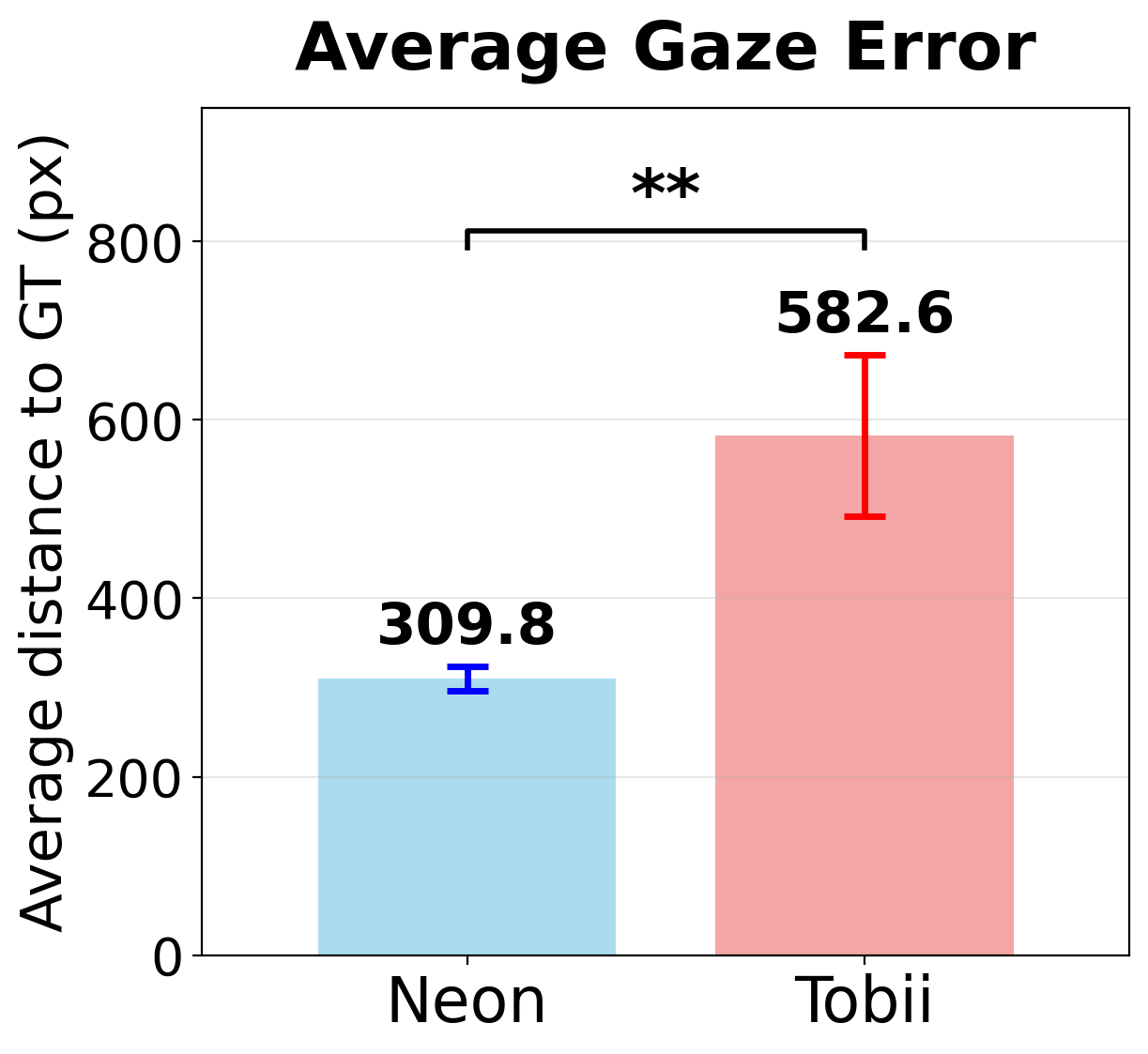}
    \caption{Average gaze error of tracking moving targets. ** indicates $p < 0.01$, * indicates $p < 0.05$.}
    \label{fig:s1avgerror}
\end{figure}

Average gaze error was $309.82 \pm 13.61$\,px for Neon and $582.58 \pm 90.87$\,px for Tobii (Fig.~\ref{fig:s1avgerror}). On a Full HD (FHD) display, these errors far exceed the size of critical anatomical targets (e.g., the cystic artery spans $\sim$200\,px; Fig.~\ref{fig:screen200px}). Even under explicit instruction to follow the tooltip, estimated gaze consistently fell outside the tooltip region (Fig.~\ref{fig:tipoutside}). A paired $t$-test showed that Neon significantly outperformed Tobii ($t = -2.97$, $p = .006$). For per-participant error plots, see Appendix Fig.~\ref{fig:s1error}.

\subsubsection{Errors are systematic and position-correlated}
\begin{figure}
    \centering
    \includegraphics[width=0.8\linewidth]{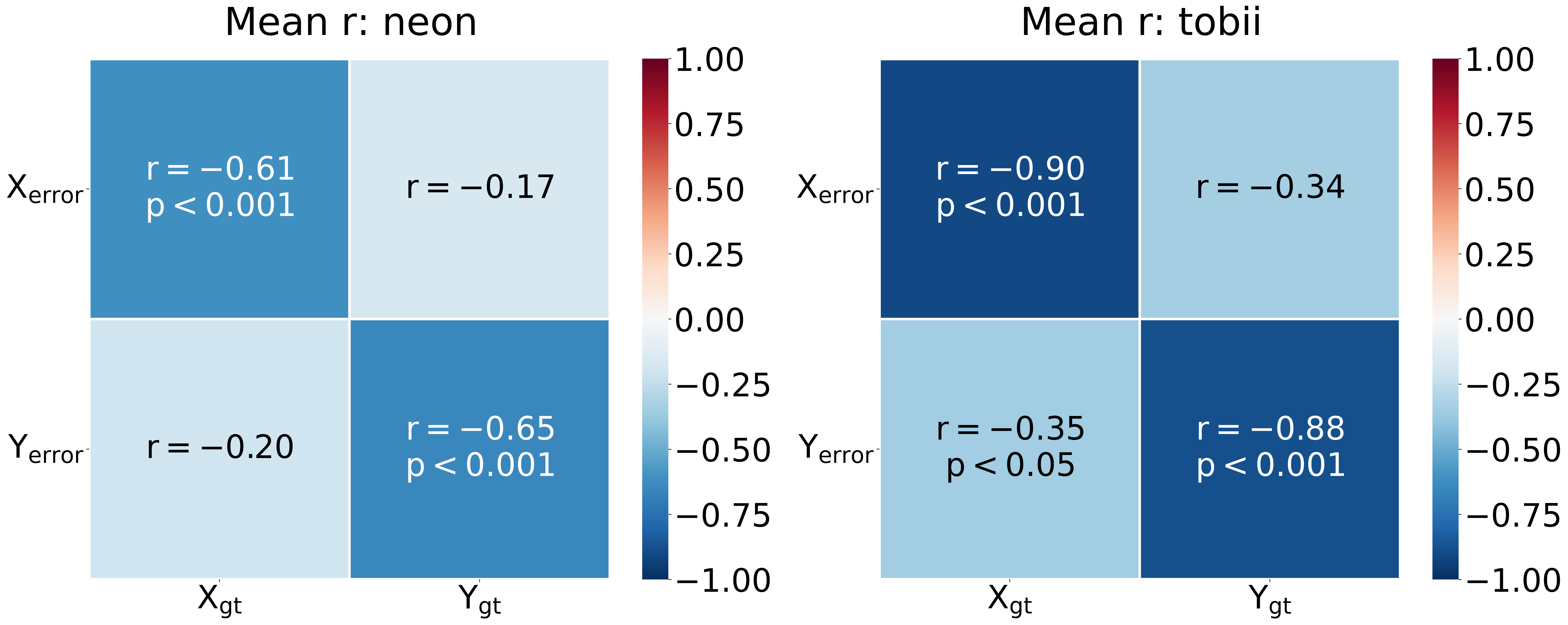}
    \caption{Correlation map between gaze errors $(X_{\text{error}}, Y_{\text{error}})$ and ground truth positions $(X_{\text{gt}}, Y_{\text{gt}})$.}
    \label{fig:s1correlation}
\end{figure}

We define gaze error per axis as the raw gaze estimate minus the ground truth tooltip position, i.e., $(X_{\text{error}}, Y_{\text{error}}) = (X_{\text{raw}}, Y_{\text{raw}}) - (X_{\text{gt}}, Y_{\text{gt}})$. Despite their magnitude, errors exhibited systematic, axis-specific patterns (Fig.~\ref{fig:s1correlation}). Pearson correlations showed that, for Neon, gaze error correlated moderately with ground truth position ($X$: $r = -0.61$; $Y$: $r = -0.65$; both $p < 0.001$). Tobii showed even stronger correlations ($X$: $r = -0.90$; $Y$: $r = -0.88$; both $p < 0.001$). Cross-axis effects were weak. This position-dependent pattern suggests errors are correctable via regression modeling (for regression plots, see Appendix Fig.~\ref{fig:s1regression}).

\subsubsection{Neon achieves higher sampling rate due to wider camera field of view}
Sampling rate is the proportion of frames in which on-screen gaze was successfully captured; higher rates capture attention more continuously and rapid eye movements more precisely, which determines the granularity of downstream gaze analysis. Neon achieved a significantly higher sampling rate ($M = 0.93$, $SD = 0.16$) than Tobii ($M = 0.67$, $SD = 0.47$; paired $t$-test, $t(29) = 3.28$, $p = .003$). This is primarily because Tobii’s narrower scene camera field of view (FoV) causes screen markers and gaze to fall outside the captured region when the head tilts, resulting in missing data (for Tobii scene camera examples, see Appendix Fig.~\ref{fig:s1tobiiscene}). Neon’s wider FoV ($1200\times1600$) maintained reliable marker visibility across head postures.

\subsubsection{Tested factors do not significantly affect Neon performance}
None of the three tested factors (lighting, head posture, content type) had a significant effect on Neon’s gaze accuracy (one-way ANOVA; head posture: $F(2,81) = 0.74$, $p = .48$; lighting: $F(2,81) = 0.29$, $p = .75$; for condition means, see Appendix~\ref{sec:s1-factors}). Tobii showed larger variability across head postures, but the difference narrowly missed significance ($F(2,81) = 1.87$, $p = .16$). Based on Neon’s superior accuracy, sampling rate, and robustness, we use Neon in all subsequent studies.

\section{SurgGaze: Implicit Calibration in Surgical Context}
Building on Study~1, we investigate \textbf{(RQ2) What calibration techniques could compensate for these errors?} We propose \textit{SurgGaze}, an implicit gaze calibration technique for the OR that models and corrects systematic eye tracking errors by using tool-tissue contact points (TTCPs) during dissection as a surrogate for the surgeon’s true gaze~\cite{zheng2021action}.
\subsection{Conceptual Model}
The premise for using TTCP draws on the established theories of eye-hand coordination in both surgical and non-surgical contexts. Classic studies on eye-hand coordination show that during object manipulation, people almost always fixate near the point where the hand contacts the object~\cite{Johansson2001EyeHand}. In surgical contexts, prior studies report that surgeons concentrate on the ongoing action, with experts tending to fixate on the target tissue while novices are more easily draw to the tool~\cite{law2004eye, Wilson2011Perceptual}. 
When the tool and target tissue physically overlap, as occurs during dissection, the surgeon’s fixation converges on the shared contact point. \textbf{We therefore define TTCP as a reliable approximation of the surgeon’s true gaze during dissection.}

Not every tool-tissue contact during dissection offers equally reliable ground truth. We purposefully selected two types of moments where we had the highest confidence in where the surgeon is fixating, based on the literature:
\begin{enumerate}
    \item \textbf{Initial contact:} when the tool first reaches the target tissue, causing the tool and anatomical target to spatially overlap. The surgeon must locate and fixate on the dissection point to guide the approach~\cite{Liu2021Developing, Li2023Using}.
    \item \textbf{Energy application:} moments when the cautery tool is actively burning or cutting tissue, which require high precision and close monitoring of each cut on the patient~\cite{bardakcioglu2015advanced}.
\end{enumerate}

These moments provide the training signal for \textit{SurgGaze} in two ways. First, the TTCP at these instants serves as a high-confidence proxy for gaze $(X_{\text{gt}}, Y_{\text{gt}})$, which is paired with the raw gaze $(X_{\text{raw}}, Y_{\text{raw}})$ from the eye tracker to form a calibration sample. Second, these samples occur inside the surgical task itself. They are therefore drawn from dynamic laparoscopic activity rather than from a separate static calibration routine. TTCPs thus operationalize the broader goal of implicit calibration: replacing sparse out-of-task calibration samples with abundant in-task samples.

Study~1 showed that gaze error varies approximately linearly with gaze location, suggesting a simple affine correction. We therefore model calibration as a 2D linear regression $f_{\theta}((X_{\text{raw}},Y_{\text{raw}})) \rightarrow (X_{\text{cal}},Y_{\text{cal}})$,
where $\theta$ denotes the model parameters. We learn $\theta$ from paired raw-ground-truth samples by minimizing the squared distance between the predicted calibrated gaze and the ground-truth fixation:
$\min_{\theta}\left\lVert f_{\theta}((X_{\text{raw}},Y_{\text{raw}}))-(X_{\text{gt}},Y_{\text{gt}})\right\rVert_2^{2}$.

\subsection{Evaluation Strategy}

The validity of \textit{SurgGaze} depends on two linked claims. First, abundant in-task calibration samples should train a better correction model than sparse out-of-task calibration samples. Second, TTCPs should provide such samples naturally in real surgery. We isolate these claims across Study~2 and Study~3.

\textbf{First, we examine the value of abundant in-task calibration samples.} Study~2 uses a controlled assigned-target task in which participants follow laparoscopic tooltips. This gives us precise reference labels for both training and testing. The 9-point baseline uses sparse static targets from an isolated calibration routine, whereas \textit{SurgGaze} uses abundant moving targets sampled from the laparoscopic-viewing task. Both methods are evaluated on the same held-out moving-target test set. Therefore, Study~2 tests the calibration-data premise, not the ecological validity of TTCPs.

\textbf{Second, we examine whether TTCPs can provide such in-task samples in authentic ORs for reducing gaze error.} Study~3 moves from assigned targets to naturally occurring surgical behavior. Here, TTCPs are evaluated as the practical mechanism for collecting in-task calibration samples during real operations. Recovering surgeons' true point-level gaze in the OR is infeasible~\cite{niehorster2024glassesvalidator}: surgeons cannot pause to annotate their exact fixation, and post-hoc pixel-level labeling is unreliable~\cite{mueller2025algorithmic}. Moreover, in OR applications, the value of eye tracking often lies in capturing ``semantic gaze''~\cite{popov2024looking}, i.e., whether gaze helps us interpret surgical behaviors such as which anatomical structure the surgeon is attending to. Accordingly, Study~3 constructs the test set as regions of interest derived from surgeons' verbal references during the procedure. If calibrated gaze more frequently falls within these regions, it suggests that TTCP-based in-task calibration is successful.

\section{Study 2: Evaluating \textit{SurgGaze} in Simulated OR}
In this study, we address \textbf{(RQ3) How effective is abundant in-task samples (implicit calibration) in simulated OR?} We use a controlled assigned-target design to isolate the effect of calibration sample source: sparse static samples from an isolated calibration routine versus abundant dynamic samples from the laparoscopic-viewing task.
\subsection{Method}
This IRB-approved study evaluates the \textit{SurgGaze} regression model under controlled conditions, comparing it against raw gaze (no calibration) and conventional 9-point explicit calibration. Beyond quantifying error reduction, we also examined practical guidelines for constructing effective calibration data. We conducted Study~2 in the same simulated OR facility as Study~1.

This design intentionally separates the calibration-data question from the TTCP-validity question. Study~2 does not claim that TTCPs are valid in the OR; instead, it tests whether the kind of data TTCPs could provide, abundant in-task samples, is useful for calibration. Study~3 then evaluates whether TTCPs can serve this role in real surgery.

\begin{figure}[t!]
    \centering
    \includegraphics[width=0.9\linewidth]{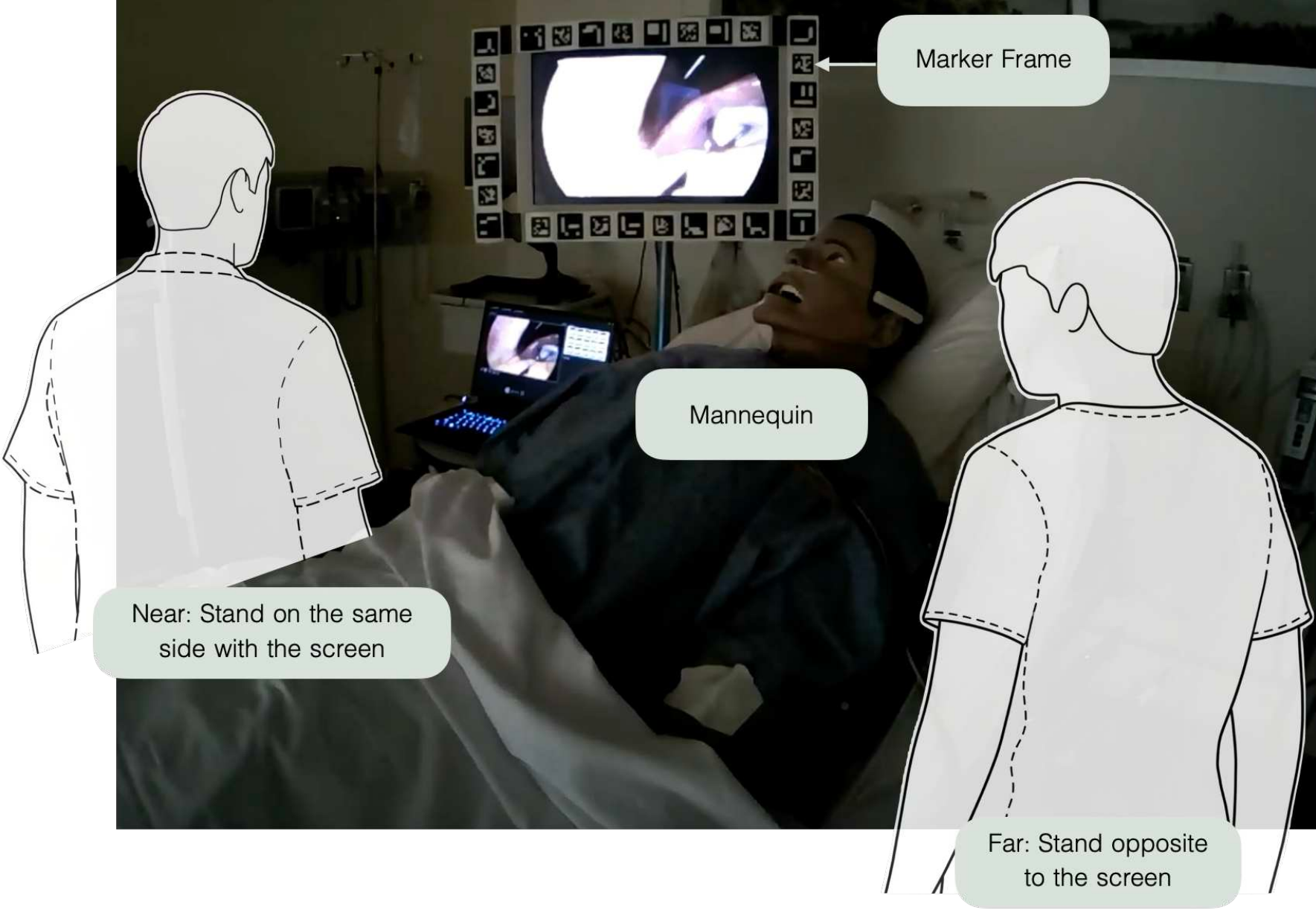}
    \caption{Experimental setup for Study 2. Participants stood either near the screen (assistant’s position) or far from the screen (primary surgeon’s position).}
    \label{fig:study2-setup}
\end{figure}

We recruited 22 participants (13F, 9M; see Table~\ref{tab:s2-participants} in Appendix) via university mailing lists, as the task only required following tooltips. Participants wore the Pupil Labs Neon eye tracker and viewed 15 minutes of lap chole video (16 segments, 25 frames per second (fps)), following a different tooltip in each segment. The assigned tooltip position served as ground truth gaze $(X_{\text{gt}}, Y_{\text{gt}})$. Participants completed the task in two viewing positions (Fig.~\ref{fig:study2-setup}): near the screen (assistant’s perspective) and far from the screen (primary surgeon’s perspective), counterbalanced across participants. Sessions lasted approximately 50 minutes; participants were compensated \$25.

We compared three conditions:
\begin{enumerate}
    \item \textbf{No Calibration:} raw gaze from Neon.
    \item \textbf{Explicit Calibration:} standard 9-point calibration, trained on assigned static fixation targets from an isolated calibration routine.
    \item \textbf{Implicit Calibration (\textit{SurgGaze}):} dense in-task calibration, trained on assigned moving tooltip targets sampled from the laparoscopic-video task. To align with the training data expected in real cases (i.e., TTCPs), we downsampled to 7{,}712 frames where the tooltip contacted target tissue.
\end{enumerate}
Each participant produced $\sim$22{,}500 gaze samples per position. The true internal gaze intent is not directly observable in either condition; therefore, assigned fixation targets serve as the reference labels for both explicit and implicit calibration. The 7{,}712 in-task calibration samples were used for training \textit{SurgGaze}; the remaining moving-target samples served as the shared held-out test set across all three conditions.

\subsection{Findings}
\subsubsection{\textit{SurgGaze} abundant in-task samples reduce error more than sparse static calibration samples}

\begin{figure*}[t]
    \centering
    \begin{subfigure}{0.48\textwidth}
        \centering
        \includegraphics[height=4cm]{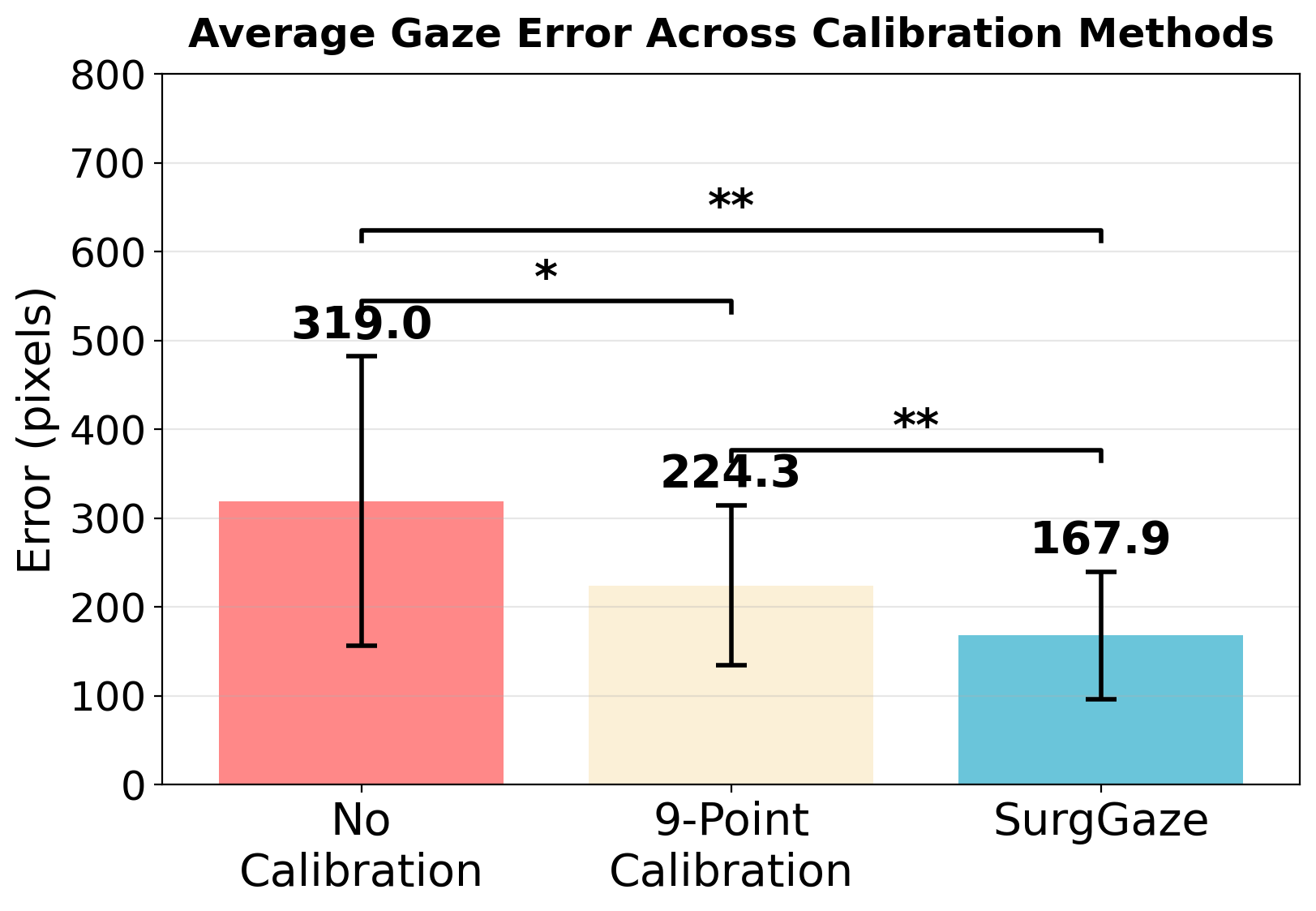}
        \caption{Average gaze error across three experimental conditions: \textit{SurgGaze} implicit calibration significantly outperforms 9-point explicit calibration, and raw gaze with no calibration.}
        \label{fig:s2avgerror}
    \end{subfigure}
    \hfill
    \begin{subfigure}{0.48\textwidth}
        \centering
        \includegraphics[height=4cm]{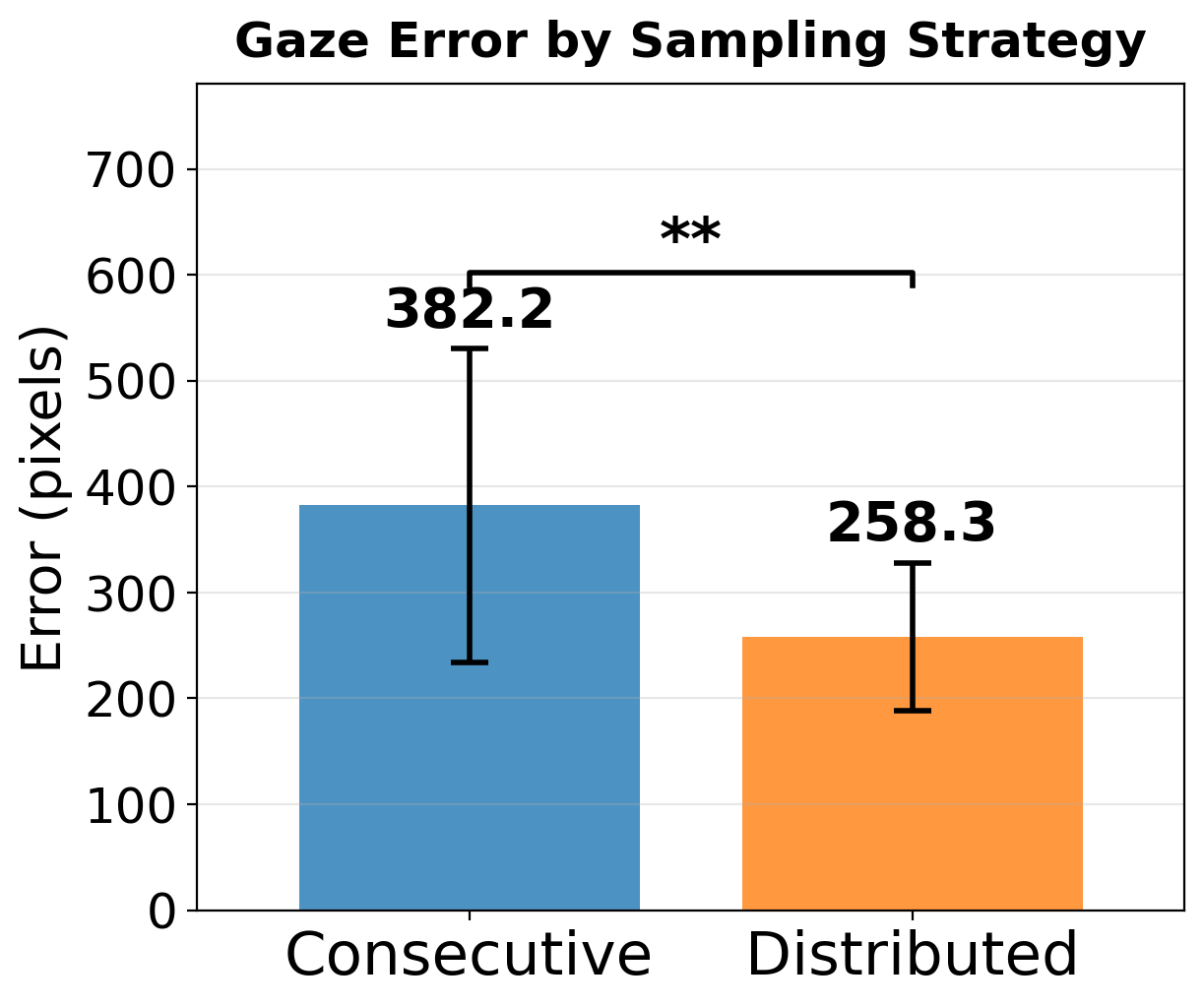}
        \caption{Average gaze error by different sampling strategies: labeling spatially distributed samples is significantly better than labeling samples from a consecutive episode.}
        \label{fig:s2avgerrorspatial}
    \end{subfigure}

    \vspace{0.6em}

    \begin{subfigure}{0.48\textwidth}
        \centering
        \includegraphics[height=4cm]{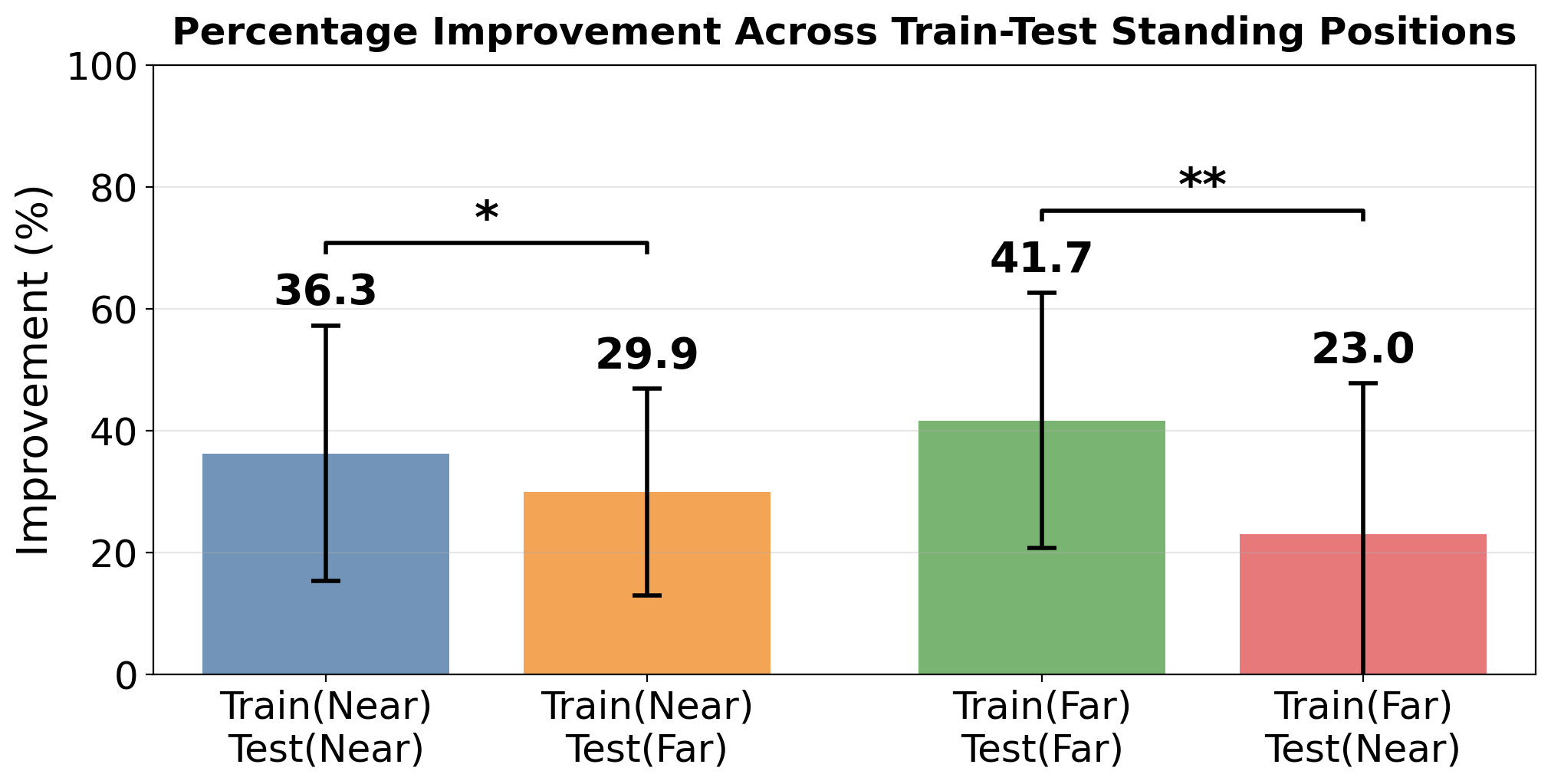}
        \caption{Average percentage improvement in gaze error reduction across the four training-testing combinations: models perform best when trained and tested/used on gaze from the same standing position.}
        \label{fig:s2standingmean}
    \end{subfigure}
\end{figure*}

We compared gaze estimation error across three experimental conditions: no calibration, explicit calibration, and implicit calibration (\textit{SurgGaze}) (Fig.~\ref{fig:s2avgerror}). A repeated-measures ANOVA revealed a significant main effect of calibration method on gaze error, $F(2,42) = 21.32$, $p < .001$, $\eta^2_G = .23$ (Greenhouse-Geisser corrected $p = 6.98 \times 10^{-5}$, $\epsilon = .56$). Average errors were highest with no calibration ($M = 319.0$, $SD = 163.0$ pixels), followed by explicit calibration ($M = 224.3$, $SD = 90.2$ pixels), and lowest with implicit calibration ($M = 167.9$, $SD = 71.7$ pixels). Paired $t$-tests showed that \textit{SurgGaze} significantly outperformed both explicit calibration ($t = -6.77$, $p < .001$, Hedges’ $g = 0.68$) and no calibration ($t = -5.57$, $p < .001$, Hedges’ $g = 1.18$). Explicit calibration also significantly reduced error compared to no calibration ($t = -3.28$, $p = .011$, Hedges’ $g = 0.71$).

\subsubsection{Calibration Point Selection: Distribution, and Sample Size}
We examined how calibration point characteristics (spatial distribution and sample size) influenced the effectiveness of the implicit calibration model.

We randomly selected samples from the training set to examine the effect of sample size. Across participants, the mean distance error decreased rapidly with the first few dozen samples, then gradually plateaued. Convergence was reached between 34--144 samples, after which additional data yielded diminishing improvements (less than 1\% over a sliding 3-window average). Convergence was defined as the point at which the error no longer improved by more than 1\% across several consecutive windows (window size = 5 samples)~\cite{Ren2022High}. Overall, \textbf{at least 150 samples are needed} to ensure stable performance across diverse users. For convergence curves and per-participant sample sizes, see Appendix Fig.~\ref{fig:s2samplesize} and Table~\ref{tab:s2convergence}.

We also compared how to select those samples. We evaluated two strategies: (1) selecting a continuous block of 150 consecutive samples from a single episode and (2) selecting points distributed across different episodes and screen regions (distributed sampling), thereby covering more positions on the screen. A paired $t$-test showed that errors were significantly lower for distributed sampling ($M = 258.3$, $SD = 69.6$ pixels) than for consecutive sampling ($M = 382.2$, $SD = 148.1$ pixels), with $t = 3.60$, $p = .002$, as shown in Fig.~\ref{fig:s2avgerrorspatial}. \textbf{Sampling discrete points distributed across the screen to maximize spatial coverage} yields better results (for per-participant comparison, see Appendix Fig.~\ref{fig:s2spatial}).

\subsubsection{Linear model with simple parameters outperforms more complex alternatives.}
We compared different regression models for calibration. The constant model performed the worst, confirming that naive estimation is insufficient. All other models, including linear regression, multilayer perceptrons (MLPs), and polynomial regressions of different degrees, achieved similar levels of performance. Importantly, adding more parameters or higher-degree polynomials did not lead to better results; in fact, simple models performed just as well. Among them, linear regression achieved a slightly lower average error compared to the others (for model comparison plot, see Appendix Fig.~\ref{fig:s2models}).

\subsubsection{Individual models should be trained from the same standing position}
We also examined whether switching the participant’s standing position affected model performance, particularly when calibration was trained on data collected on one side and applied to data from the other. Paired $t$-tests revealed that accuracy improvements were significantly higher when models were trained and tested on the same side. As shown in Fig.~\ref{fig:s2standing} and Fig.~\ref{fig:s2standingmean}, training on the near side (the whole training set with 7,712 points) and testing on the near side (near$\rightarrow$near, $M = 36.3\%$, $SD = 21.0\%$) outperformed training on the near side and testing on the far side (near$\rightarrow$far, $M = 29.9\%$, $SD = 17.0\%$), $t = 2.85$, $p = .010$. Likewise, training on the far side and testing on the far side (far$\rightarrow$far, $M = 41.7\%$, $SD = 21.0\%$) outperformed training on the far side and testing on the near side (far$\rightarrow$near, $M = 23.0\%$, $SD = 24.8\%$), $t = 4.25$, $p < .001$. These results indicate that implicit calibration models degrade in performance when applied across standing positions, and work best when trained and tested on the same.

\section{Study 3: Evaluating \textit{SurgGaze} in Authentic OR}
Having established the value of abundant in-task calibration samples under controlled assigned-target conditions, we next ask whether real surgery naturally provides such samples. To address \textbf{(RQ4) How effective is \textit{SurgGaze}, which uses TTCPs as abundant in-task calibration samples, in authentic OR settings?}, Study~3 evaluates whether TTCPs extracted from real surgeries provide a reliable calibration signal, and whether the resulting calibration improves interpretation of which object the surgeon is attending to.

We followed the model-training guidelines from Study~2: (1) at least 150 spatially distributed calibration samples, (2) a linear regression model, and (3) a separate model for each standing position.

\subsection{Participants and Procedure}
\begin{figure}[t!]
    \centering
    \includegraphics[width=0.8\columnwidth]{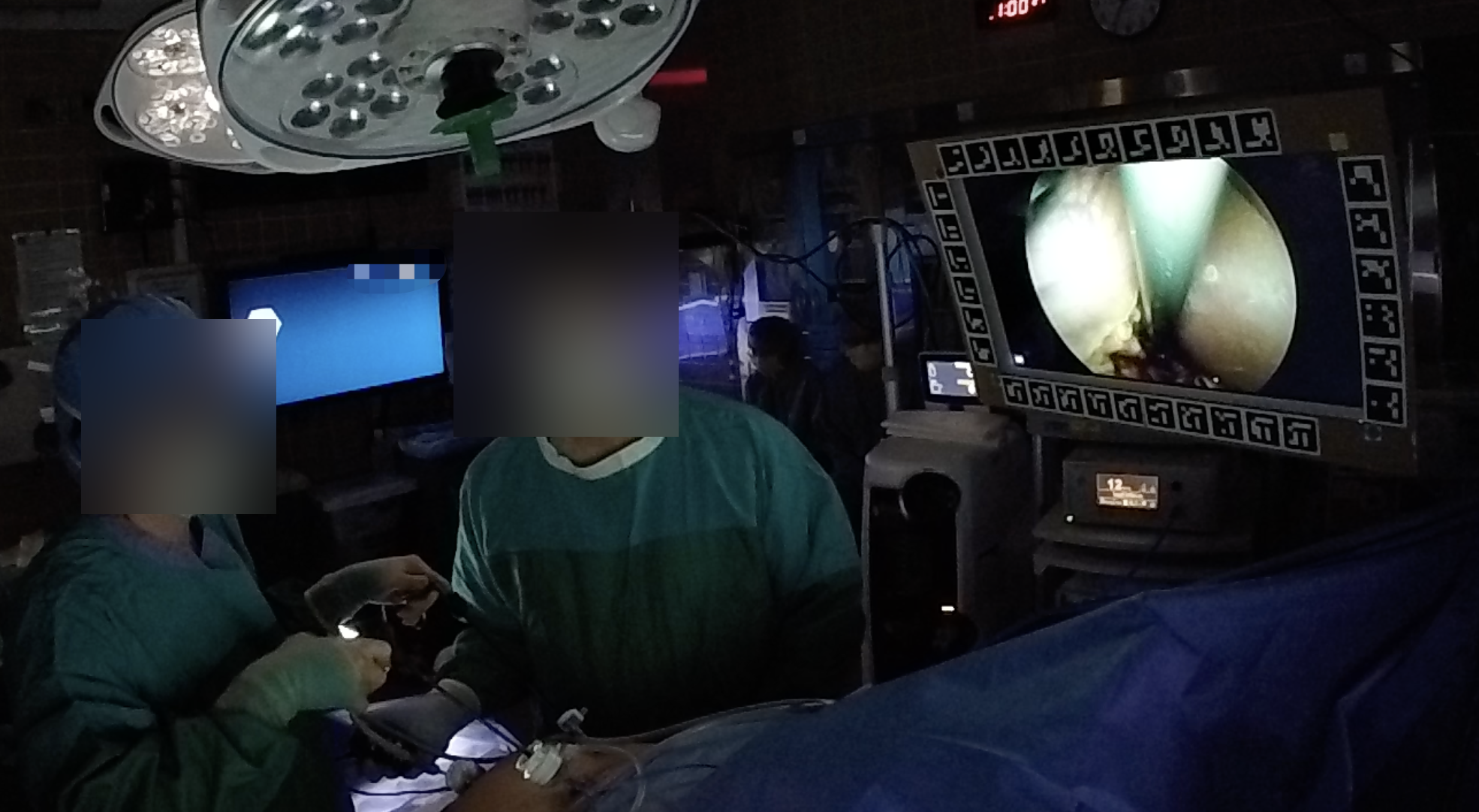}
    \caption{Authentic OR setup for Study~3.}
    \label{fig:study3or}
\end{figure}

This IRB-approved study included five surgeons across four live lap chole cases at a major U.S. teaching hospital (3 attendings, 2 residents; 4M, 1F; all White; see Table in Appendix). Surgeons were recruited based on the schedule of eligible cases. We obtained consent from both the surgeons and the patients. Study~3 used the same screen size, marker frames for screen recognition, and Neon eye tracker as Study~2, but was conducted in an authentic OR with two laparoscopic displays on the left and right sides of the patient (Fig.~\ref{fig:study3or}). Prior to each surgery, researchers attached marker frames to the laparoscopic displays and assisted the operating surgeon in wearing the Neon eye tracker before scrubbing in. After the operation, the laparoscopic video was exported for analysis. Participant s3-2 wore the glasses only during the beginning of surgery, resulting in less data. Because real-time eye tracking studies in authentic ORs are especially difficult to conduct, recruiting five surgeons represented a substantial practical achievement, and each surgery yields large volumes of gaze data.

\subsection{Data Processing}
We describe how the raw data was preprocessed, and how training and test sets were constructed.
\subsubsection{Preprocessing}
After exporting the gaze data from the Neon system, we first synchronized the Neon front-facing camera video with the laparoscopic video stream through the following steps:
\begin{enumerate}
    \item Extract raw gaze points and front-camera video from the Pupil Labs Neon system.  
    \item Synchronize Neon recordings with the laparoscopic machine video. Visual landmarks for alignment include turning the laparoscopic machine on/off (screen transition from black to blue) or the moment when the laparoscopic camera is inserted. When these events appear in both the eye tracker’s front-camera recording and the laparoscopic machine feed, we use them as synchronization anchors.
    \item From the Neon front-camera view, review video content to determine the operator’s role and standing position:  
    \begin{itemize}
        \item Assistant (left side): tools typically enter from the left, with more grabbing and retracting actions.  
        \item Main surgeon (right side): tools typically enter from the right, with more cutting and dissecting actions.  
    \end{itemize}
    \item Collect training data separately for each role/side, as tool dynamics differ between assistant and main surgeon.  
\end{enumerate}
We then extracted moments in which the surgeon served as the primary operator, focusing on segments of high engagement, either critical dissection events (training set) or moments of concentrated attention and verbal explanation (test set).  

\subsubsection{Training Set}
A critical step of using \textit{SurgGaze} is to identify the calibration sample -- frames in which the surgeon's gaze point on the screen is known with high certainty. 
In the current implementation, we used a guided manual approach in which a medical expert (a medical student at M2 level or above who has completed a surgery intern rotation) followed a routine to select and annotate frames. It took the medical expert about 15 minutes to label and annotate \textasciitilde150 frames. 
In this section, we briefly introduce the steps. We provide a detailed codebook in Section~\ref{sec:codebook} (Appendix) to help others replicate the protocol. 

\begin{figure*}
    \centering
    \includegraphics[width=0.8\linewidth]{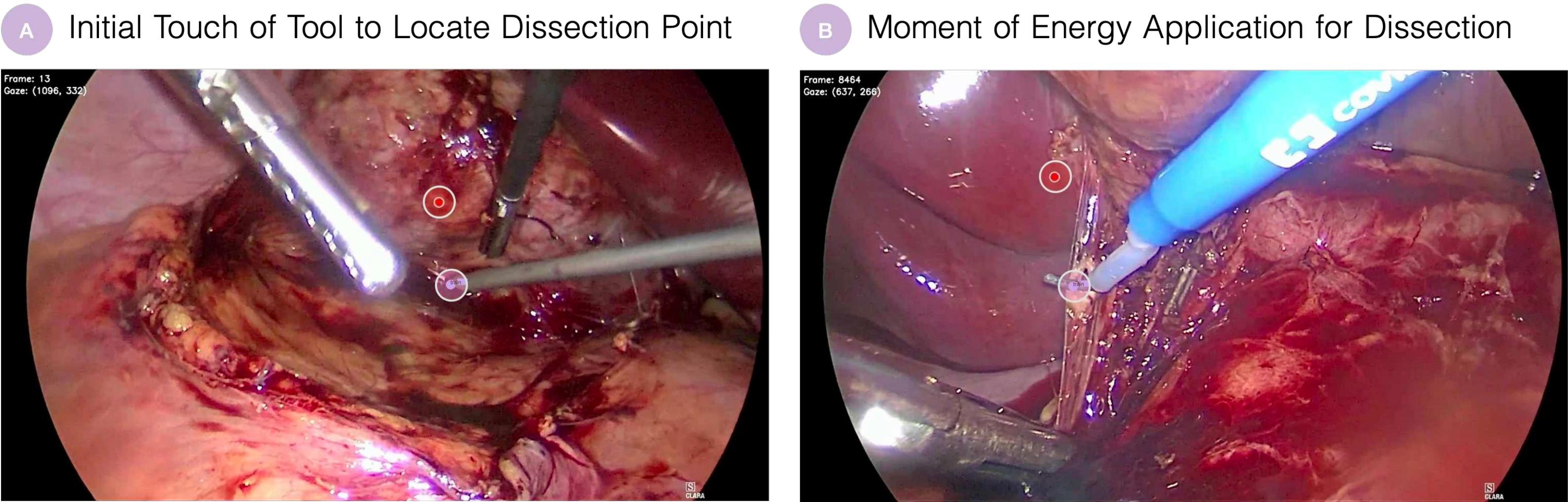}
    \caption{Examples of training samples used for implicit calibration, illustrating key surgical moments: initial tool contact to locate the dissection point (A) and energy application for tissue dissection (B).}
    \label{fig:s3trainlabel}
\end{figure*}

Calibration samples are identified based on tool-tissue contact, which provides the most reliable implicit gaze targets. The medical expert first identifies the dissection period, which usually lasts \textasciitilde20 minutes. They then look for frames that the tool contacts tissue, and among those frames, carefully select frames that satisfy the following criteria, as shown in Fig.\ref{fig:s3trainlabel}: 
\begin{enumerate}
    \item The initial touch of the single-ended tool, when the surgeon locates the dissection point. This moment captures the convergence of the surgeon’s two typical fixation targets during actions, the tool and the tissue, at the point where they intersect.
    \item The moment of application of energy for tissue dissection, when tool-tissue contact occurs, and the surgeon must focus precisely on the burn site. This moment represents a high-stakes phase of eye-hand coordination that requires increased attention and close monitoring.
\end{enumerate}
If yes, they will label the contact point between tooltip and tissue as the ground truth gaze point. The medical expert used a labeling interface\footnote{\url{https://github.com/CVHub520/X-AnyLabeling}} for this task. 

We excluded phases where gaze is not reliably anchored, such as approach/navigation, hole creation, surrounding-tissue cleaning, and tool withdrawal, to prevent noise in the training set. This careful selection was validated directly by s3-3, who confirmed: \textit{“Yes, this is smart, we really look at those places (tooltip-tissue contact point during dissection).”}

Study 2 suggests that approximately 150 annotated frames are sufficient for calibration, so that the medical expert annotated \textasciitilde150 frames for each of the 5 cases. During the process, the expert was prompted to distribute their sample selections across the screen to ensure spatial diversity. The researchers reviewed the spatial distribution by plotting the distribution of labels. The medical expert added additional samples as needed to compensate for underrepresented regions. Although our current protocol is manual, it is amenable to automation to improve scalability and ensure broader spatial coverage across the display.

\subsubsection{Test Set}
During the surgery operation, we asked the participating surgeons to verbalize or point on the screen where they were looking. We then reinstalled the specific frames where the surgeons were describing where they were looking. We then defined ROIs based on surgeons' verbal references.
To avoid overlap between training and testing data, we excluded all moments involving tools that were heavily used during calibration-sample selection. This ensures that testing samples are not inadvertently similar to the training set. Our goal is to validate that a model trained only on TTCPs can generalize its corrections to other visually attended targets, such as the gallbladder, fat, holes, nodes, etc.

\begin{figure*}
    \centering
    \includegraphics[width=\linewidth]{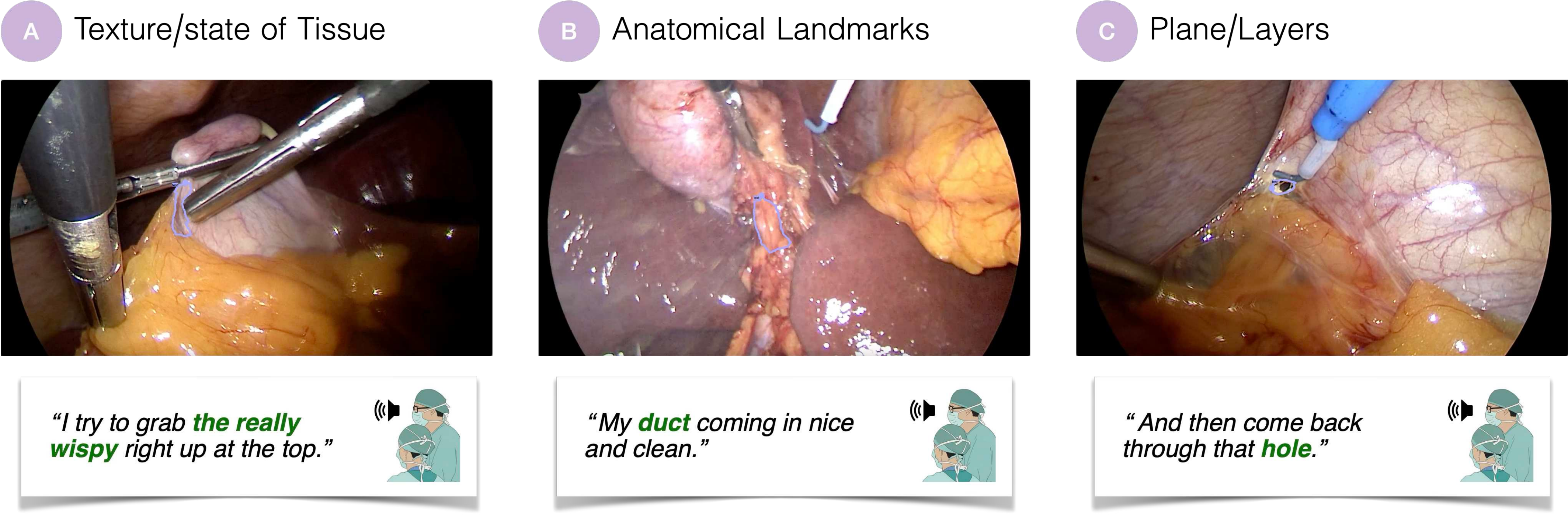}
    \caption{Examples of Study~3 test samples derived from expert-labeled communication, with segmented ROIs.}
    \label{fig:s3testlabel}
\end{figure*}

Researchers first reviewed the transcripts of the surgical recordings and identified sentences in which surgeons explicitly referenced their visual attention (e.g., \textit{“I try to grab the really wispy right up at the top.”}) as shown in Fig.\ref{fig:s3testlabel}. Each reference was then mapped back to the corresponding object in the synchronized laparoscopic video. A medical expert first annotated the region of interest in the initial frame of each sentence span using a labeling interface\footnote{\url{https://github.com/CVHub520/X-AnyLabeling}}
 integrated with Segment Anything Model 2 (SAM2)~\cite{Ravi2024SAM}. Researchers then ran SAM2 to propagate these annotations and produce segmentation masks for the referenced structures throughout the full duration of the verbal reference. In cases where the verbal reference was brief (typically under 10 frames at 25~fps), we treated this duration as the span of attentional focus. SAM2 was then applied across these short temporal windows to propagate segmentations frame by frame, ensuring consistent representation of the referenced target. This process produced a set of semantically grounded ROIs that reflect the surgeons’ attention region.
\subsubsection{Metrics}
When constructing the test set based on surgeons' verbal references, the targets were often small regions. The annotations defined ROIs rather than single points. To evaluate the gaze calibration results, we measured the percentage of calibrated gaze that fell into the ROI. 
This allowed us to assess not only pixel-level accuracy but also whether calibrated gaze could reliably capture the semantic target of the surgeon’s visual attention. 

\subsection{Quantitative Results}
\begin{figure*}[t!]
    \centering
    \includegraphics[width=0.8\linewidth]{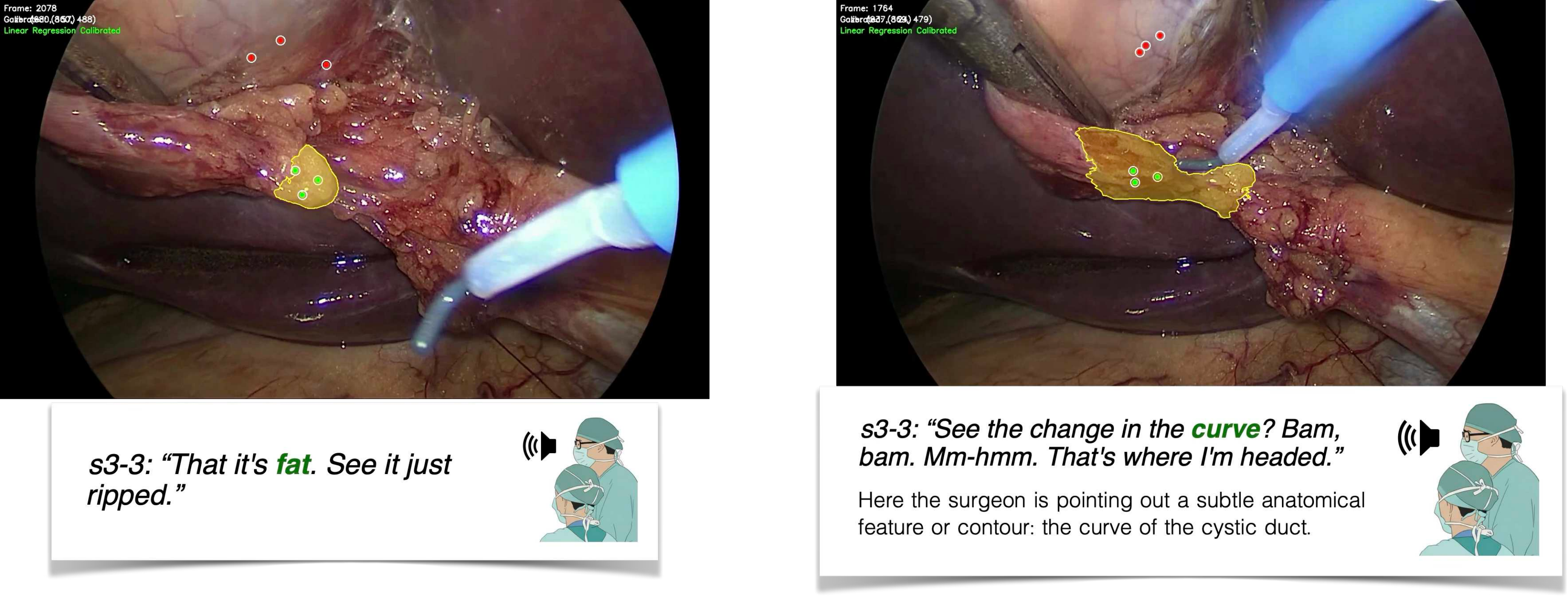}
    \caption{Examples where \textit{SurgGaze}-calibrated gaze aligns with surgeons’ verbal references. Green highlights mark the annotated ROI (semantic landmark: ``fat''); red dots indicate raw gaze and green dots indicate calibrated gaze.}
    \label{fig:s3success}
\end{figure*}

Across five participants (totaling 441 training samples and 582 test samples), raw gaze data rarely aligned with the annotated visual context: on average, only 11.34\% of raw gaze points fell within the intended regions of interest. As summarized in Tab.~\ref{tab:s3result}, after applying calibration, alignment improved dramatically, with 85.75\% of gaze points falling into the regions of interest, which is a 8 times improvement over raw gaze on average. Fig.~\ref{fig:s3success} illustrates several cases where surgeons explicitly referenced anatomical landmarks (e.g., “That it’s \textit{fat}. See it just ripped.”; “See the change in the \textit{curve}? That’s where I’m headed.”),  In those cases, gaze calibration allowed the surgeon’s visual focus to be aligned with the referenced region.

\begin{table}[h]
\begin{tabular}{rcccc}
\hline
\makecell[l]{Participant\\ID} &
\makecell[c]{\# Training\\Samples} &
\makecell[c]{\# Test\\Samples} &
\makecell[c]{Raw Gaze\\in Context} &
\makecell[c]{Calibrated Gaze\\in Context} \\
\hline
s3-1 & 152 & 190 & 20.53\% & 96.84\% \\
s3-2 & 58  & 20  & 10.00\% & 90.00\% \\
s3-3 & 151 & 266 & 0.00\%  & 87.97\% \\
s3-4 & 153 & 145 & 13.10\% & 79.31\% \\
s3-5 & 157 & 151 & 3.97\%  & 73.51\% \\
SUM  & 671 & 772 & 11.34\% & 85.75\% \\
\hline
\end{tabular}

\caption{With \textit{SurgGaze}, calibrated gaze are 8 times more likely to fall in the regions of interest. (Study 3)
}
\label{tab:s3result}
\end{table}

\subsection{Qualitative Examples}
Beyond the quantitative evaluation, we examined unlabeled segments to assess whether calibrated gaze produced predictions that were plausible and meaningful. These segments are different from the test set, for which we have high confidence of the surgeon's gaze point. The following examples demonstrate that \textit{SurgGaze}-calibrated gaze is aligned with existing theories of surgical expertise (e.g., anticipatory gaze, target-focused attention) and has the potential to reveal new attention patterns that were previously unobservable due to gaze inaccuracy.

\subsubsection{Accurate localization of intended grasping area on two-ended tools}
\begin{figure}[t!]
    \centering
    \includegraphics[width=\linewidth]{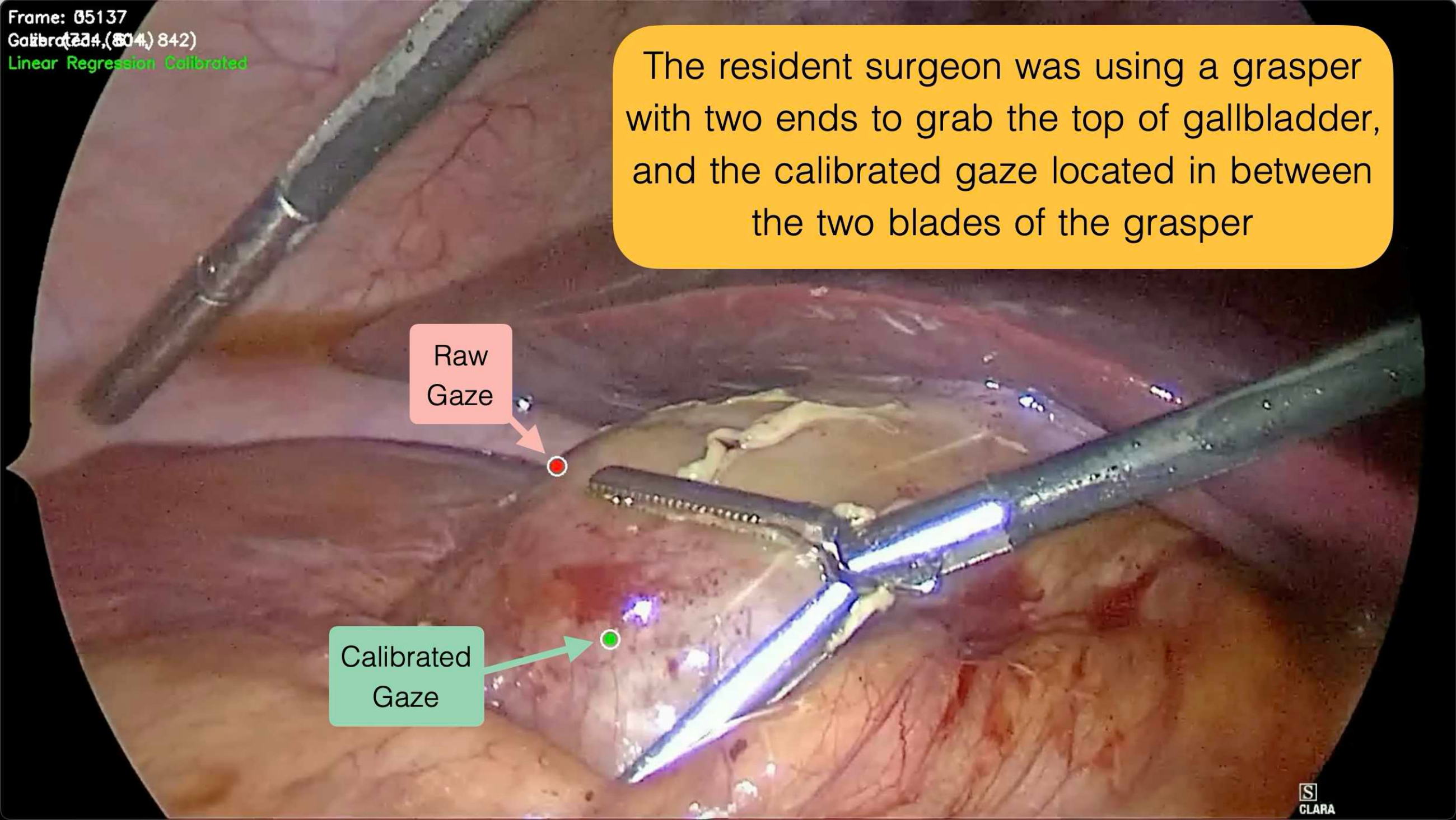}
    \caption{Example of calibrated gaze when the scene has an active two-ended tool, such as a grasper.}
    \label{fig:s3betweengrasper}
\end{figure}

We intentionally avoided using two-ended instruments such as graspers in the calibration sample, since there is ambiguity of the fixation of the operating surgeon. Surgeon’s gaze may fall anywhere along or between the jaws. We observed that the calibrated gaze on scenes that have an active grasper consistently localized between the two tips, on top of the tissue being grasped (Fig.~\ref{fig:s3betweengrasper}). In contrast, the raw gaze often fell outside the target region, highlighting the value of calibration in capturing meaningful points of surgical attention.

\subsubsection{Prediction of the intended action area before contact}

\begin{figure*}[t!]
    \centering
    \includegraphics[width=\linewidth]{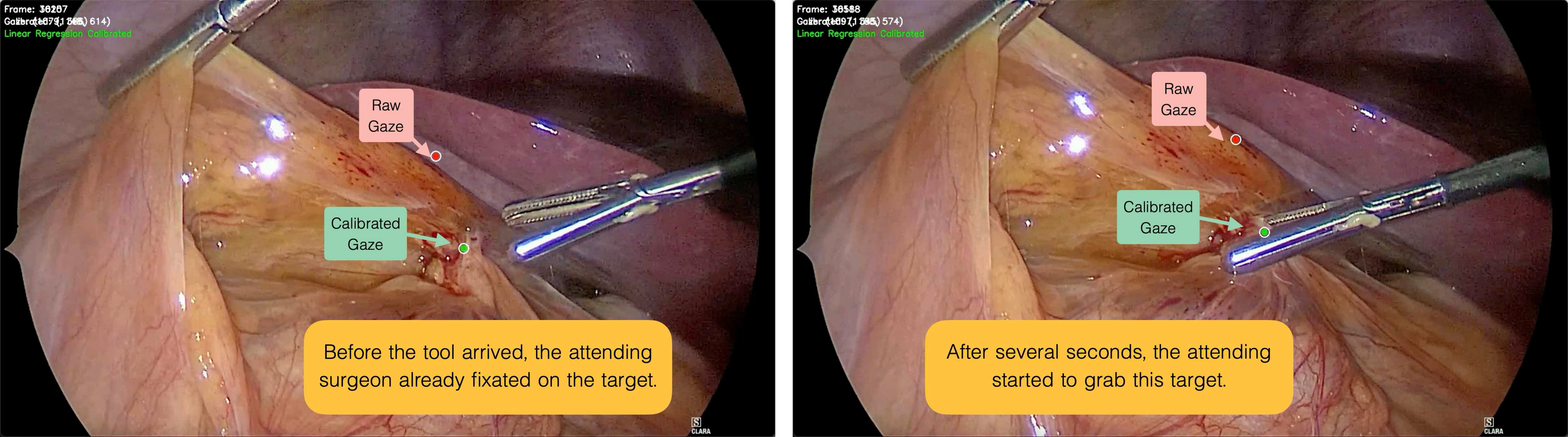}
    \caption{Calibrated gaze was on the intended action area before the grasper reached the target tissue (left); several seconds later, the instrument moved to grasp the action area (where the calibrated gaze fell into)  (right).}
    \label{fig:s3predicttarget}
\end{figure*}

Fig.~\ref{fig:s3predicttarget} presents an example where the calibrated gaze aligns with the surgeon's intended action, e.g., the calibrated gaze was on the intended action area before the grasper reached the target tissue. A few seconds later, the instrument moved to grasp the action area (where the calibrated gaze fell into). This shows that the calibrated gaze can be used to predict surgeons' next-step actions. 

In contrast, the raw gaze remained on the liver, which was not the operative target at this stage. 

\subsubsection{Calibrated gaze aligned with surgeons' descriptions when they used the tool as a pointer}
\begin{figure*}[t!]
\centering
\includegraphics[width=\linewidth]{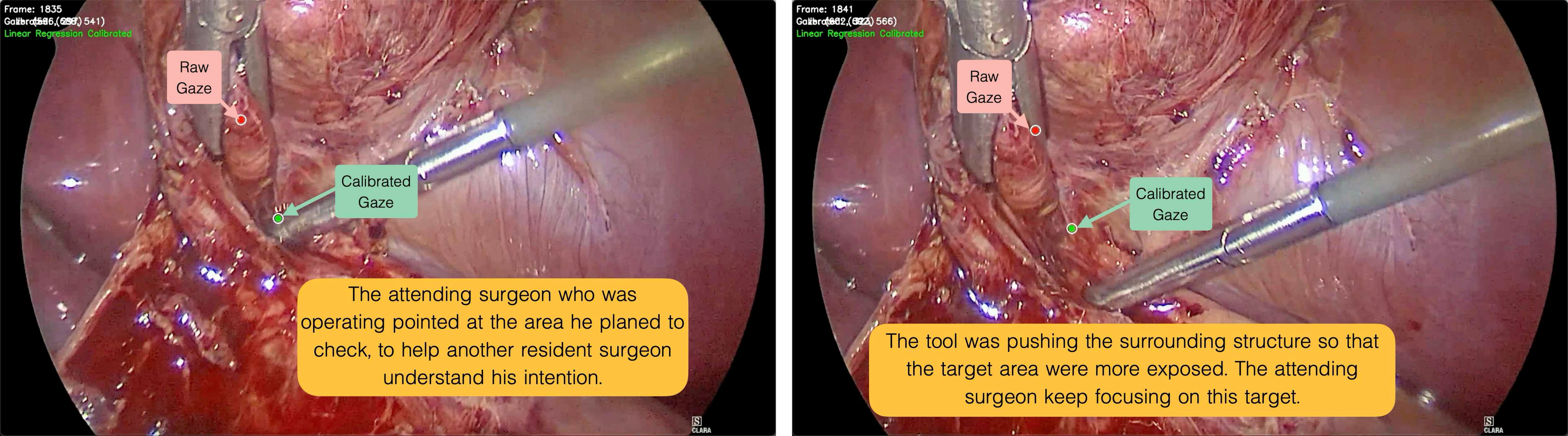}
\caption{Calibrated gaze tracked the grasper-indicated target (not the tool), even without verbal cues, and stayed on the true target when the tool later retracted.}
\label{fig:s3alignwithinstruction}
\end{figure*}

Beyond dissection, instruments were sometimes used as pointers to communicate surgical attention. In these cases, surgeons pointed with the tip of a tool to highlight structures for the team. The calibrated gaze was closely aligned with the indicated region, even when the surgeon did not verbally specify the target (e.g., saying only “a little over here” while pointing with the grasper; Fig.~\ref{fig:s3alignwithinstruction}). Notably, the gaze did not shift as the tool later pushed surrounding tissue to provide better exposure, indicating that the tool was no longer serving as a cursor but as a retractor. This shows that calibrated gaze can capture implicit attentional cues expressed through tool-mediated gestures. By contrast, the raw gaze fell between the jaws of the left grasper, which was retracting the gallbladder for exposure, rather than aligning with the surgeon’s true attentional target.

\subsubsection{Gaze fixation maintained on target after tool retraction}
\begin{figure*}[t!]
    \centering
    \includegraphics[width=\linewidth]{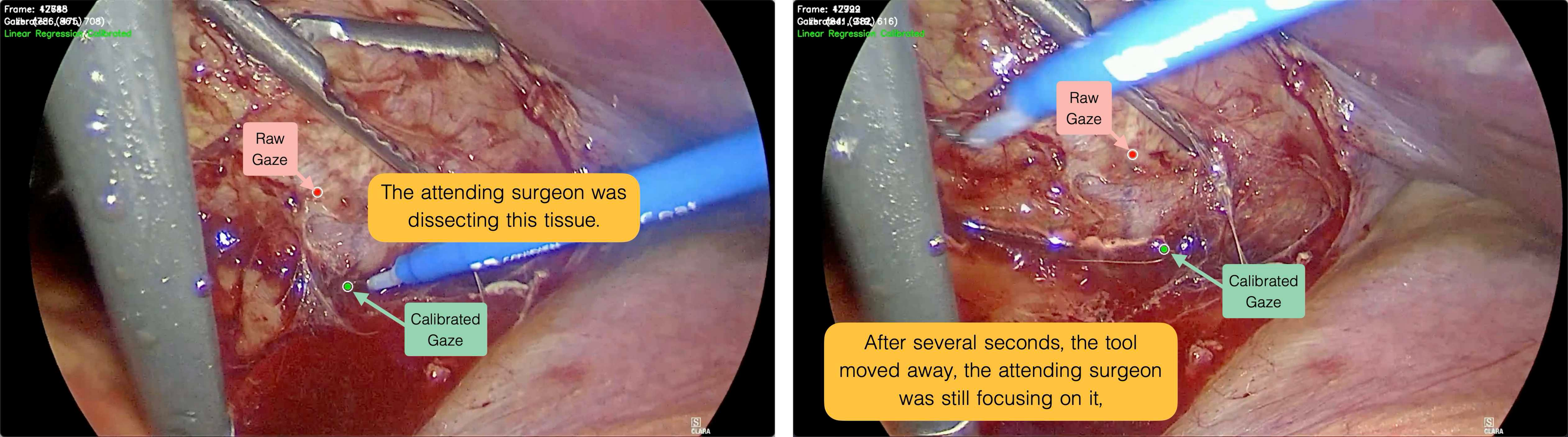}
    \caption{Calibrated gaze stayed on the tissue target after retraction, while raw gaze always drifted, reflecting expert-level attentional stability.}
    \label{fig:s3ontargetafterdissection}
\end{figure*}

Even after the instrument was retracted, the calibrated gaze remained fixated on the original target area, consistent with expert annotations (Fig.~\ref{fig:s3ontargetafterdissection}). This behavior aligns with prior work~\cite{khan2012analysis}, which showed that expert surgeons tend to maintain stable fixations on their targets rather than shifting prematurely. In contrast, the raw gaze drifted upward onto another part of the gallbladder, missing the actual dissection site.

\subsubsection{Consistency with finger pointing cues}
\begin{figure*}[t!]
    \centering
    \includegraphics[width=\linewidth]{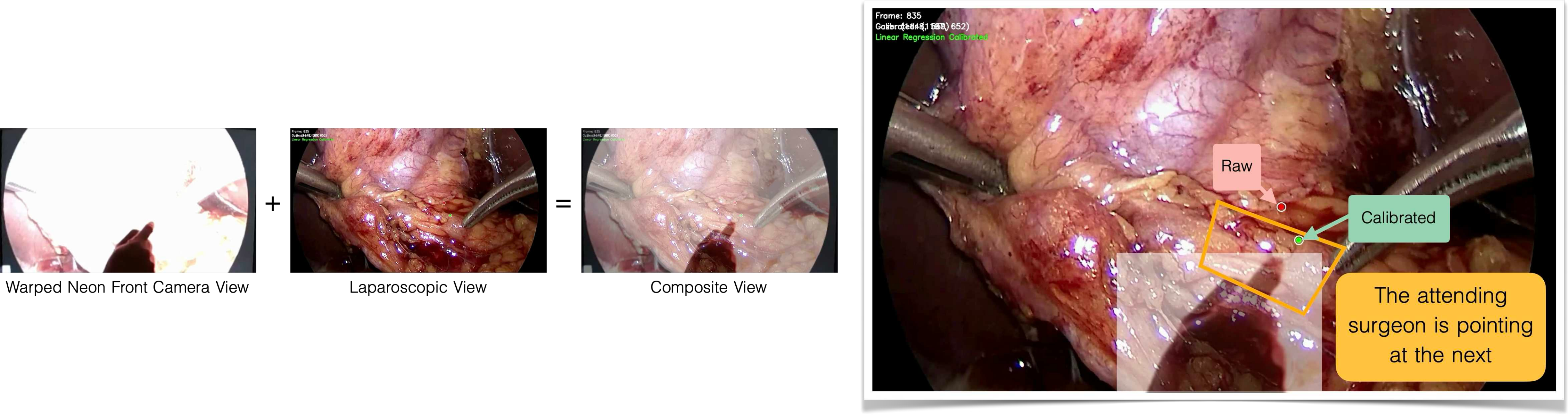}
    \caption{Calibrated gaze more closely aligns with where teammates point on the screen with their fingers.}
    \label{fig:s3fingerpointing}
\end{figure*}

We also observed moments of instruction and discussion in which team members pointed directly at the laparoscopic monitor to indicate structures for dissection. By extracting the screen area and overlaying gaze points, we found strong alignment between calibrated gaze and these referential gestures. Specifically, as shown in Fig.~\ref{fig:s3fingerpointing}, the calibrated gaze consistently fell just above the pointing finger, whereas the raw gaze often drifted in the wrong direction. This indicates that when other team members pointed to the monitor, the predicted gaze aligned with those gestures, reflecting how surgeons attend to and engage with collaborative visual references.

\section{Discussion}
 
In this work, we introduce and validate \textit{SurgGaze}, an implicit calibration approach that leverages structured behaviors during surgery, specifically tool-tissue contact points, to calibrate surgeons' gaze in authentic ORs. The central contribution of \textit{SurgGaze} is not only removing the need for an isolated calibration routine. More importantly, it reframes calibration as a data-source problem: effective gaze correction benefits from calibration samples that are abundant and collected within the activity where gaze will be analyzed. Study~2 shows the value of this shift under controlled assigned-target conditions, and Study~3 shows that TTCPs can operationalize it in real surgical practice. In five live surgery cases, \textit{SurgGaze}-calibrated gaze was 8 times more likely to fall within regions that surgeons were attending to compared with raw gaze.

The OR evaluation study further demonstrates that \textit{SurgGaze}-calibrated gaze could help uncover behavioral patterns that signal surgical expertise, such as 1) anticipatory gaze shifts to the next dissection target before tool movement; 2) precise fixation on tissue positioned between grasping tool jaws; and 3) sustained attention on surgical targets rather than instrument tracking. These gaze-informed patterns may enable finer-grained analyses of surgical attention, including gaze path plots and detection of cognitive lapses, and enhance our understanding of how attending and resident surgeons coordinate during operations.

\subsection{Generalizability Beyond Lap Chole}
Although this study was contextualized in Lap Chole, the approach could be generalized to other laparoscopic procedures (e.g., appendectomy, hernia repair). The display settings and video capture conditions are consistent across laparoscopic surgeries, and the high-stakes dissection moments used as training data occur frequently in a wide range of procedures. This suggests that the proposed calibration framework could generalize beyond Lap Chole to support \textit{SurgGaze} in diverse laparoscopic contexts. The key requirement for \textit{SurgGaze} is the availability of single-tip tool-tissue interactions. Although this paper uses the L-hook, any single-tip laparoscopic instrument provides the same one-point contact needed for reliable calibration, and such tools appear in nearly all laparoscopic procedures. Common examples include cautery instruments (hook or spatula)~\cite{Anis2019Comparative}, which are used for cutting, dissecting, and coagulating and are present in the vast majority of cases; suction-irrigators~\cite{Lee2020Irrigation}, which clear the field and manage bleeding and are frequently used in laparoscopy and universally in robotic surgery; and graspers or dissectors used with closed jaws to create small openings in tissue~\cite{Khan2023Force-adjustable}, representing a fundamental technical skill. All of these instruments offer clear, single-point tool-tissue contacts suitable for generating calibration samples.

\subsection{Potential Sources of Training Data}
In practical deployments, additional events could also be leveraged as training data. Examples include finger pointing at the monitor, tool tips used as cursors to indicate structures, or explicit verbal references to anatomical landmarks during discussion. These naturally occurring referential cues can serve as implicit calibration signals, reducing reliance on dedicated calibration routines.  
The confidence of different types of training data remains an open question for future work. In this paper, we focus on validating the feasibility of implicit calibration, without yet examining the relative reliability of each training data source.

\subsection{Automation and Future Interfaces}
\label{sec:automation}
The current annotation process during the training phase could be further automated. For example, after identifying dissection periods, frames could be sent to a vision-language model to determine whether the tool tip is contacting tissue. Advanced segmentation methods could also generate a more comprehensive computational understanding of the surgical scene, supporting reliable identification of tool-tissue interactions. Future interfaces could integrate AI modules to automatically detect candidate dissection moments, track the distribution of previously labeled training data, and recommend additional frames for annotation.

\subsection{Implicit Gaze Calibration with Structured Behavior Signals}
This work provides empirical evidence on deploying mobile eye tracking in a high-stakes environment: surgery, extending prior work on implicit gaze calibration leveraging touchscreen interactions \cite{Cai2025GazeSwipe:}. It also validates a broader conceptual approach: implicit calibration based on heterogeneous and structured behavioral signals can effectively improve gaze calibration in the wild, where people act on the physical world through a mediated visual interface. 
This opens up new avenues for applying implicit calibration in other real-world tasks and domains, such as teleoperation, telestration, and construction involving camera-assisted equipment. In these settings, users similarly coordinate vision and action around task-relevant targets while relying on screen-based visual mediation. Our findings therefore suggest a broader design opportunity for HCI: to develop calibration methods that are tailored to the behavioral structure of a domain, enabling more robust gaze estimation in the wild without disrupting ongoing work.

\subsection{Scope and Scalability}
Because gaze error patterns depend on how the eye tracker sits on the wearer's face and on the geometric relationship between the wearer and the display, \textit{SurgGaze} trains a separate calibration model per surgeon, per wear session, and per standing position. These are the same constraints that apply to explicit calibration, which must similarly be repeated each time the device is worn or the user moves. In practice, laparoscopic surgery involves only two standing positions (primary surgeon and assistant), and annotating 150 calibration frames takes approximately 15 minutes per position. As TTCP detection becomes automated (\S\ref{sec:automation}), even this effort could be further reduced.

\subsection{Limitation}
In Study~3, marker-based frame recognition occasionally failed under lighting changes or occlusion, and very small targets ($<$50\,px) remained difficult to resolve even after calibration. \textit{SurgGaze} is most reliable in central screen regions during active dissection. We also found fewer calibration samples for assistant-side surgeons, which may reduce robustness; future work could incorporate proxy ground truth from other channels. Our studies used different participant populations: Studies~1--2 used non-expert students, whereas Study~3 involved attending surgeons with lower demographic diversity, reflecting broader workforce imbalances.\footnote{\url{https://datausa.io/profile/soc/surgeons}} We evaluated only two devices (Tobii Pro Glasses~2 and Pupil Labs Neon); future work should test additional and newer eye trackers.

\section{Conclusion}
In this paper, we present \textit{SurgGaze}, an implicit calibration method for eye tracking in laparoscopic surgery. By leveraging natural eye-hand-tool coordination, specifically tool-tissue contact during dissection, \textit{SurgGaze} avoids a separate calibration routine while shifting calibration from sparse out-of-task samples to abundant in-task samples. Study~2 shows that such in-task samples improve pixel-level calibration compared with conventional 9-point calibration, and Study~3 shows that TTCPs can provide this calibration resource in authentic ORs. Beyond improved accuracy, \textit{SurgGaze} captures meaningful signals of expertise such as anticipatory gaze and target-focused attention, enabling higher-level analyses of technical and non-technical practices in surgery.

\begin{acks}
This work was supported by the National Science Foundation under Grant No. 2406218, ``Multimodal Techniques to Enhance Intra- and Post-operative Learning and Coordination between Attending and Resident Surgeons.''
\end{acks}
\bibliographystyle{ACM-Reference-Format}
\bibliography{sample-base}
\appendix
\section{Appendix for Study 1}
\subsection{Participant Demographic}
Table~\ref{tab:s1participants} summarizes the demographics of the 14 participants in Study~1.
\begin{table}[htbp]
\centering
\small
\begin{tabular}{ccccc}
\hline
ID & Gender & Ethnicity & Vision & Major \\
\hline
 s1-1 & Female &  Asian East& No glasses  & Engineering\\
 s1-2 & Male & Asian East& No glasses  & Engineering\\
 s1-3 & Female & White & Contacts  & Medicine\\
 s1-4 & Female & Asian South& Contacts  & Medicine\\
 s1-5 & Male & Asian South& Glasses-NoContact  & Engineering\\
 s1-6 & Male & Asian East& Glasses-NoContact  & Engineering\\
 s1-7 & Female & Asian East& Contacts  & Engineering\\
 s1-8 & Male & Asian South& Contacts  & Engineering\\
 s1-9 & Male & White & Glasses-NoContact  & Medicine\\
 s1-10 & Male & Asian East& Contacts  & Engineering\\
 s1-11 & Male & White & Contacts  & Engineering\\
 s1-12 & Female & Asian South& Contacts  & Engineering\\
 s1-13 & Male & Middle Eastern Asian & Glasses-NoContact  & Engineering\\
 s1-14 & Female & African American & Glasses-NoContact  & Engineering\\
\hline
\end{tabular}
\caption{Study 1 participant demographics.}
\label{tab:s1participants}
\end{table}

\subsection{Workflow}
\begin{figure*}[htbp]
    \centering
    \includegraphics[width=\linewidth]{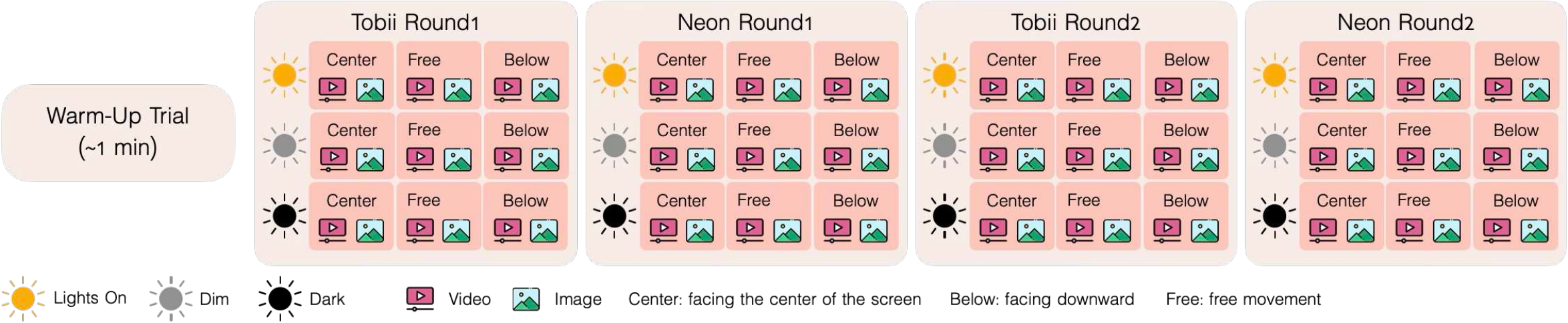}
    \caption{Procedure for Study~1. Each participant first completed a $\sim$1~min warm-up trial, followed by 72 experimental trials divided into two rounds of 36. The design was fully crossed: 2 eye trackers (Tobii, Neon) $\times$ 2 content types (video, image) $\times$ 3 lighting conditions (bright, medium, dark) $\times$ 3 head positions (center, free, below).}
    \label{fig:s1procedure}
\end{figure*}

Fig.~\ref{fig:s1procedure} illustrates the full procedure for Study~1.
\subsection{Gaze error of Individual Participants}
\begin{figure*}[htbp]
    \centering
    \includegraphics[width=\linewidth]{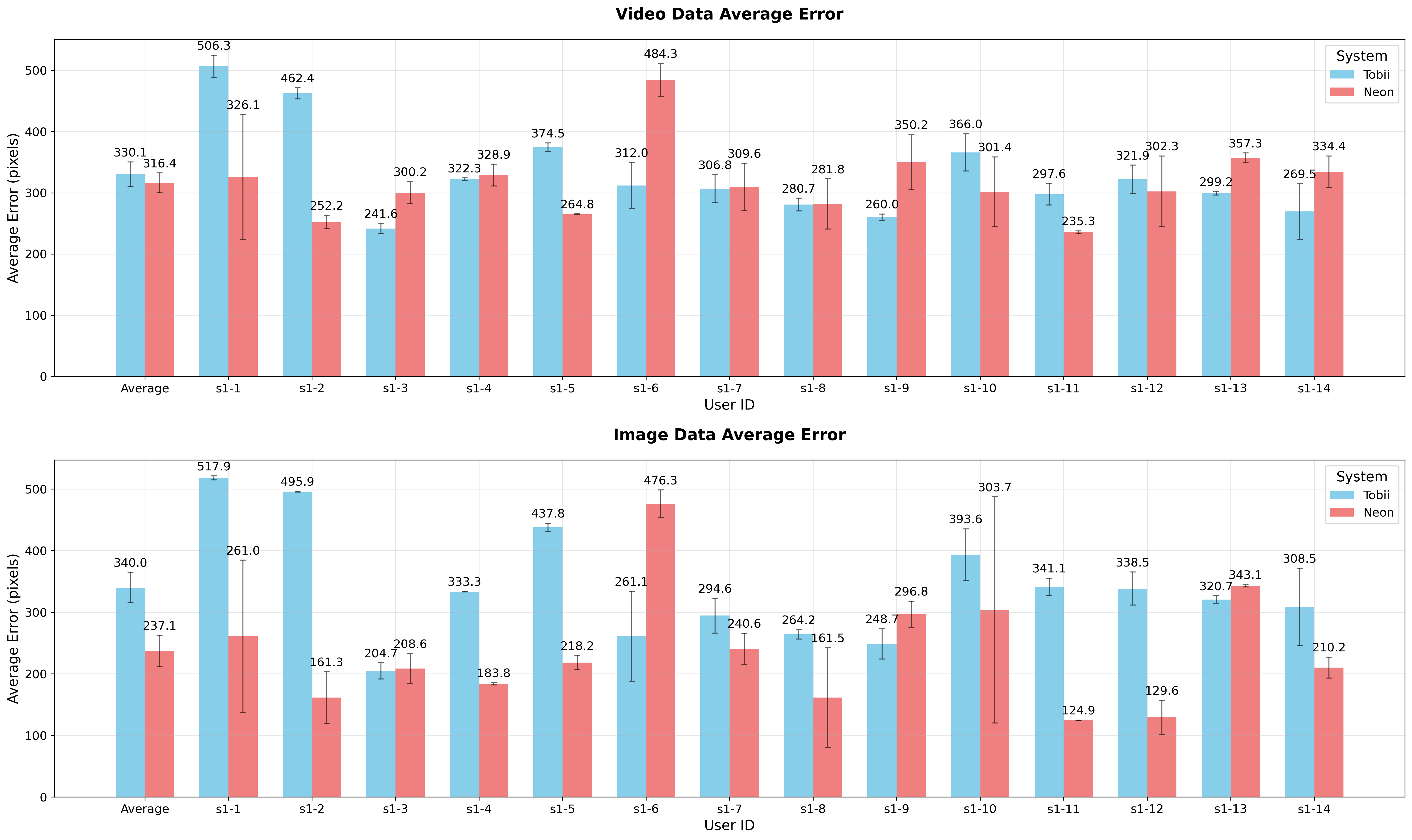}
    \caption{Average gaze error (pixels) across 14 participants in Study 1, comparing Tobii Pro Glasses 2 (blue) and Pupil Labs Neon (red) under facing center and lights on condition (most favorable condition for sampling rate). Results are shown separately for (top)  video clips and (bottom) images. Errors were computed as the distance between raw gaze points and the ground truth tooltip center.}
    \label{fig:s1error}
\end{figure*}

Fig.~\ref{fig:s1error} shows the per-participant gaze error for both devices.
\subsection{Sample Rate in Study 1}
Statistical analysis (Fig.~\ref{fig:s1samplevsframe} in Appendix) further showed that Tobii’s sampling rate is moderately correlated $(r = 0.325, p < 0.001)$ with marker detection rate, suggesting that incomplete screen capture directly impacts tracking reliability.
\begin{figure}[htbp]
    \centering
    \includegraphics[width=0.8\linewidth]{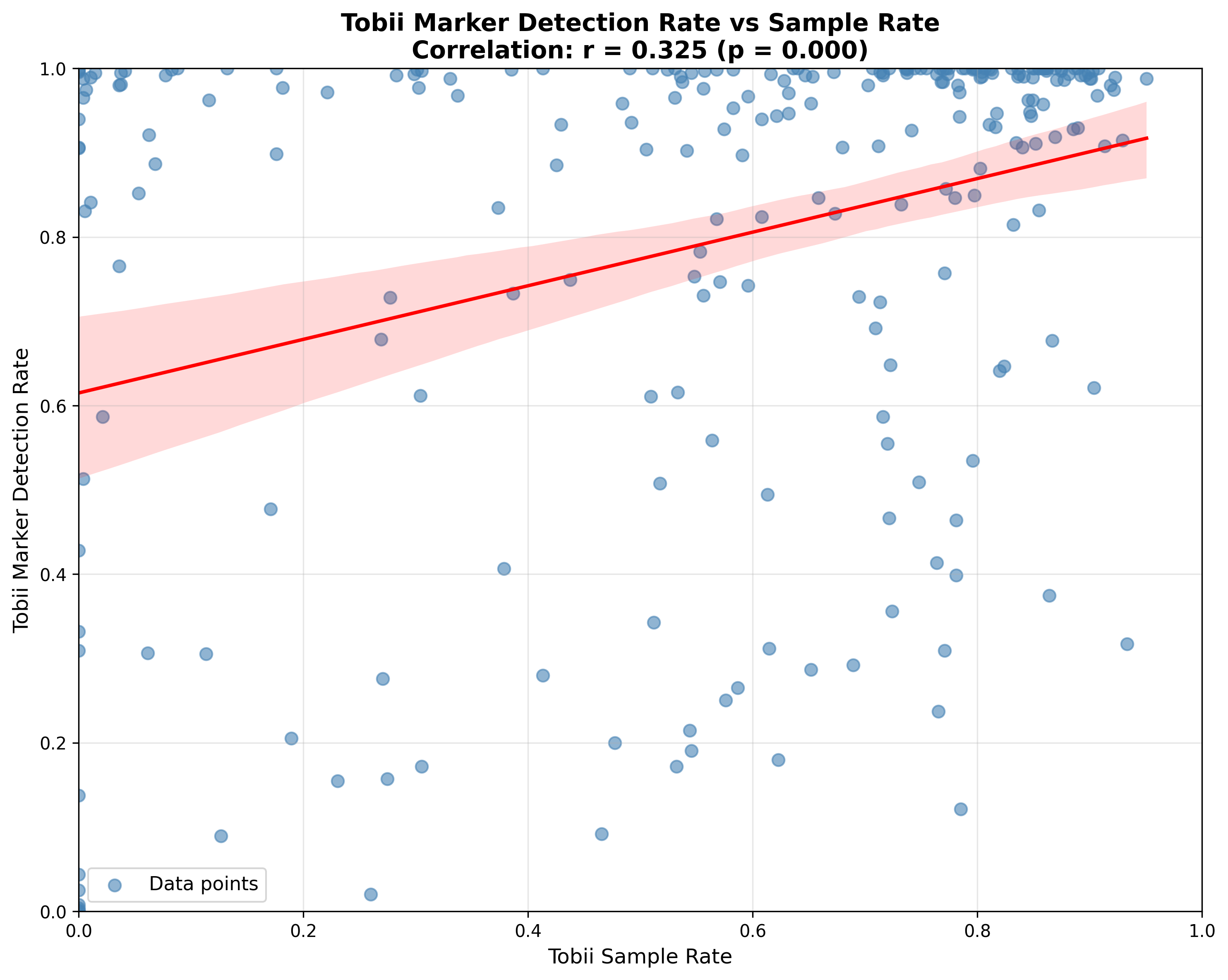}
    \caption{Relationship between Tobii sample rate and marker detection rate across Study 1 trials. Each point represents a participant-condition pair. The red line indicates the best-fit regression, with the shaded region showing the 95\% confidence interval. A moderate positive correlation was observed $(r = 0.325, p < 0.001)$, suggesting that higher sample rates are associated with more reliable marker detection.}
    \label{fig:s1samplevsframe}
\end{figure}
\subsection{Tobii Scene Camera Field of View}
\begin{figure}[htbp]
    \centering
    \includegraphics[width=0.5\linewidth]{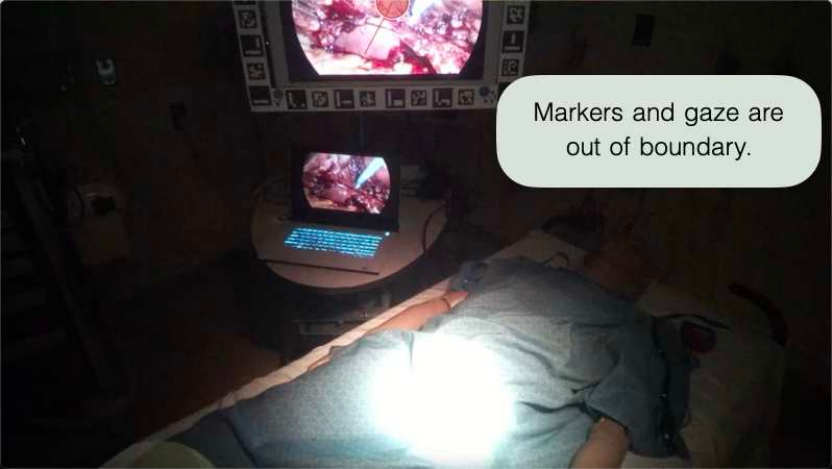}
    \caption{When participants tilt their head downward, Tobii often loses track of the markers and the screen, pushing them out of the view.}
    \label{fig:s1tobiiscene}
\end{figure}

Tobii's greater sensitivity to head-facing position is primarily due to the field of view (FoV) of its front-facing scene camera. Tobii's camera is limited to a $1920\times1080$ resolution and often crops out the upper portion of screen markers. As a result, both markers and surgeons' gaze frequently fall outside the captured region (Fig.~\ref{fig:s1tobiiscene}).

\subsection{Head Pose and Lighting Effects on Accuracy}
\label{sec:s1-factors}
For Neon, head-aiming position had no significant effect: Center ($309.82 \pm 13.61$ px), Below ($323.24 \pm 15.35$ px), and Free ($297.22 \pm 13.83$ px) showed comparable accuracy ($F(2,81) = 0.74$, $p = 0.481$). Tobii displayed larger variability across head positions: Center ($582.58 \pm 90.87$ px), Below ($926.2 \pm 192.50$ px), Free ($750.443 \pm 93.7$ px), but this difference narrowly missed significance ($F(2,81) = 1.87$, $p = 0.16$).

Lighting conditions similarly produced no significant effect for either device. Neon performance remained stable across Lights On ($309.82 \pm 13.61$ px), Dark ($323.37 \pm 16.25$ px), and Dim ($325.36 \pm 16.45$ px) ($F(2,81) = 0.29$, $p = 0.75$). Tobii showed no lighting differences either: Lights On ($582.58 \pm 90.87$ px), Dark ($586.56 \pm 91.86$ px), and Dim ($597.68 \pm 94.88$ px) ($F(2,81) = 0.007$, $p = 0.99$).

\subsection{Linear Regression in Study 1}
\begin{figure*}[htbp]
    \centering
    \includegraphics[width=\linewidth]{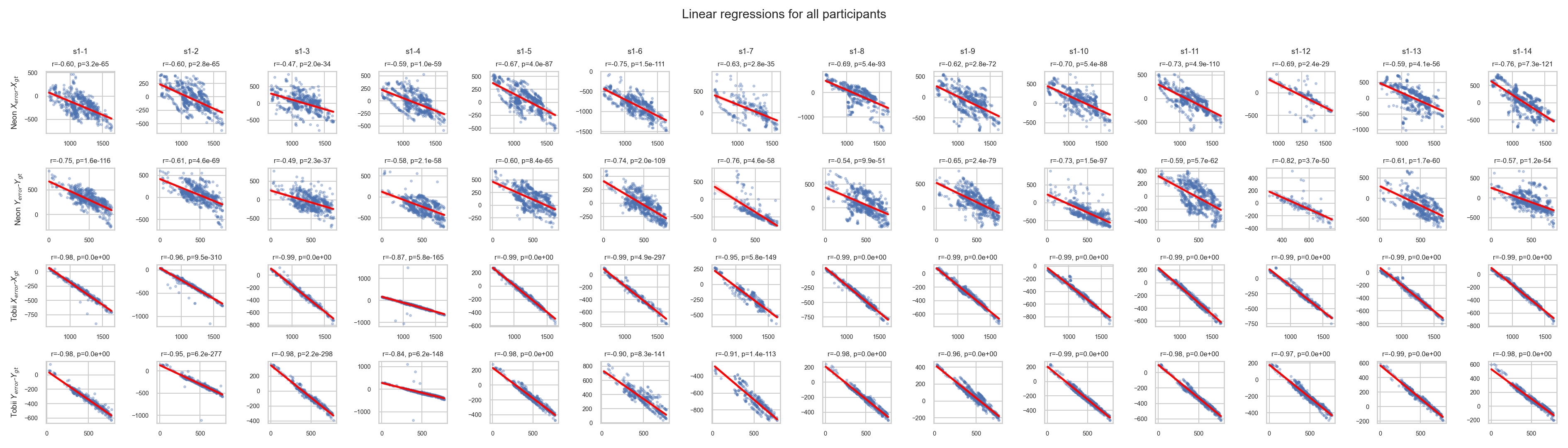}
    \caption{Linear regression of gaze errors $(X_{\text{error}}, Y_{\text{error}})$ and ground truth positions $X_{\text{gt}}, Y_{\text{gt}}$. }
    \label{fig:s1regression}
\end{figure*}

Fig.~\ref{fig:s1regression} shows the linear regression between gaze errors and ground truth positions for both devices.
\section{Appendix for Study 2}
\subsection{Participant Demographic}
Table~\ref{tab:s2-participants} summarizes the demographics of the 22 participants in Study~2.
\begin{table}[htbp]
\small
\begin{tabular}{rcccc}
\hline
ID    & Gender & Ethnicity          & Vision  &Major \\ \hline
s2-1  & Female  & African American              & No glasses & Engineering     \\
s2-2  & Male   & Asian East         & No glasses & Engineering     \\
s2-3  & Female & Asian East         & No glasses & Medicine     \\
s2-4  & Male   & Asian East         & Contact lens & Engineering   \\
s2-5  & Female & White              & No glasses  & Engineering    \\
s2-6  & Female & White              & No glasses  & Engineering    \\
s2-7  & Male   & White              & No glasses  & Engineering    \\
s2-8  & Female & White              & No glasses & Engineering     \\
s2-9  & Female & Asian East         & No glasses  & Engineering    \\
s2-10 & Male   & Asian South        & Neon lens  & Engineering     \\
s2-11 & Male   & Asian East         & Neon lens   & Engineering    \\
s2-12 & Female & Asian East         & No glasses  & Engineering    \\
s2-13 & Male   & White              & Contact lens  & Engineering  \\
s2-14 & Male   & Asian East         & No glasses   & Engineering   \\
s2-15 & Female & Asian East         & No glasses   & Engineering   \\
s2-16 & Female & Hispanic or Latino & No glasses  & Medicine    \\
s2-17 & Female & Asian East         & No glasses   & Engineering   \\
s2-18 & Female & Hispanic or Latino & No glasses   & Engineering   \\
s2-19 & Male   & Asian East         & Neon lens   & Engineering    \\
s2-20 & Male   & White              & Neon lens  & Engineering     \\
s2-21 & Male   & Hispanic or Latino & No glasses   & Engineering   \\
s2-22 & Female & White              & Contact lens   & Engineering \\ \hline
\end{tabular}
\caption{Study 2 participant demographics.}
\label{tab:s2-participants}
\end{table}

\subsection{Workflow}
\begin{figure}[htbp]
    \centering
    \includegraphics[width=\columnwidth]{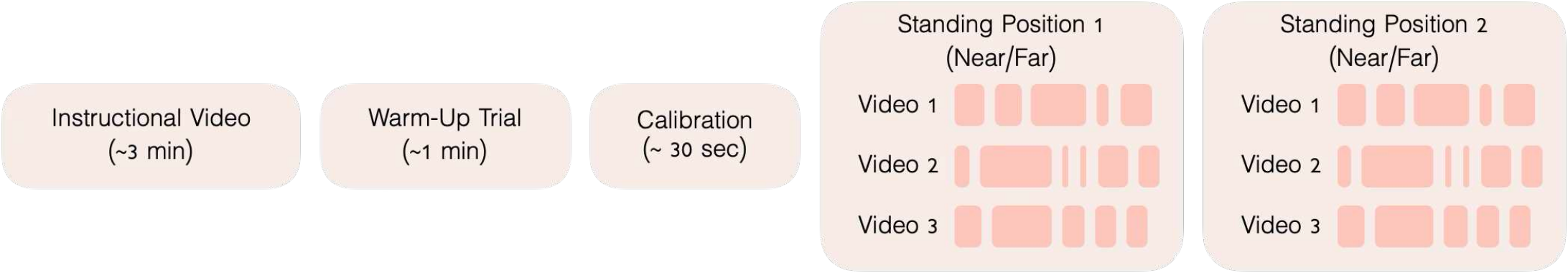}
    \caption{Study 2 procedure. Participants completed preparation steps, then viewed lap chole videos in two standing positions, following a single tool tip per segment.}
    \label{fig:study2-procedure}
\end{figure}

Fig.~\ref{fig:study2-procedure} illustrates the full procedure for Study~2.
\subsection{Per-Participant Gaze Error}
\begin{figure*}[htbp]
    \centering
    \includegraphics[width=\linewidth]{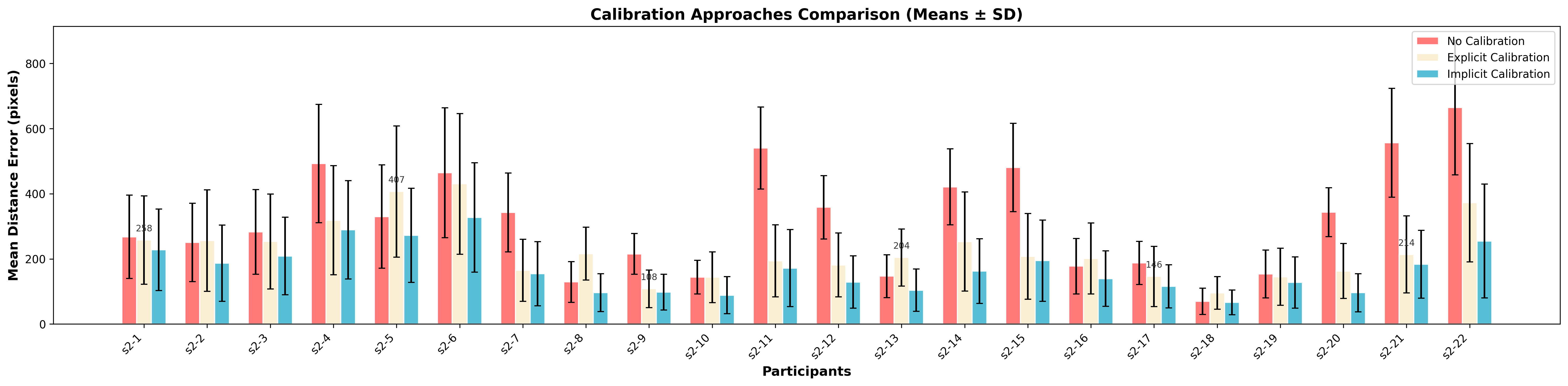}
    \caption{Comparison of mean distance error across calibration strategies for 22 participants in Study 2. \textit{SurgGaze} (Implicit Calibration) has the lowest average errors among the three conditions.}
    \label{fig:calibration-accuracy}
\end{figure*}

Fig.~\ref{fig:calibration-accuracy} compares per-participant gaze error across the three calibration conditions.
Across participants, the dominant pattern followed raw > explicit > implicit calibration performance (e.g., s2-1, 3, 4, 6, 7, 9, 10, 11, 12, 14, 15, 17, 19, 20, 21, 22). However, for participants s2-2, 5, 8, 13, 16, and 18, explicit calibration even worsened accuracy. For s2-8, 13, 16, and 18, the raw gaze error was already small, leaving minimal room for correction; in these cases, suboptimal models like explicit calibration can mislead the gaze, and implicit calibration could only make marginal adjustments. In contrast, s2-2 and s2-5 appeared to experience cognitive slips during the explicit calibration task, resulting in noisy training samples that degraded calibration quality.

\subsection{Standing position effects on calibration}
\begin{figure*}[htbp]
    \centering
    \includegraphics[width=\linewidth]{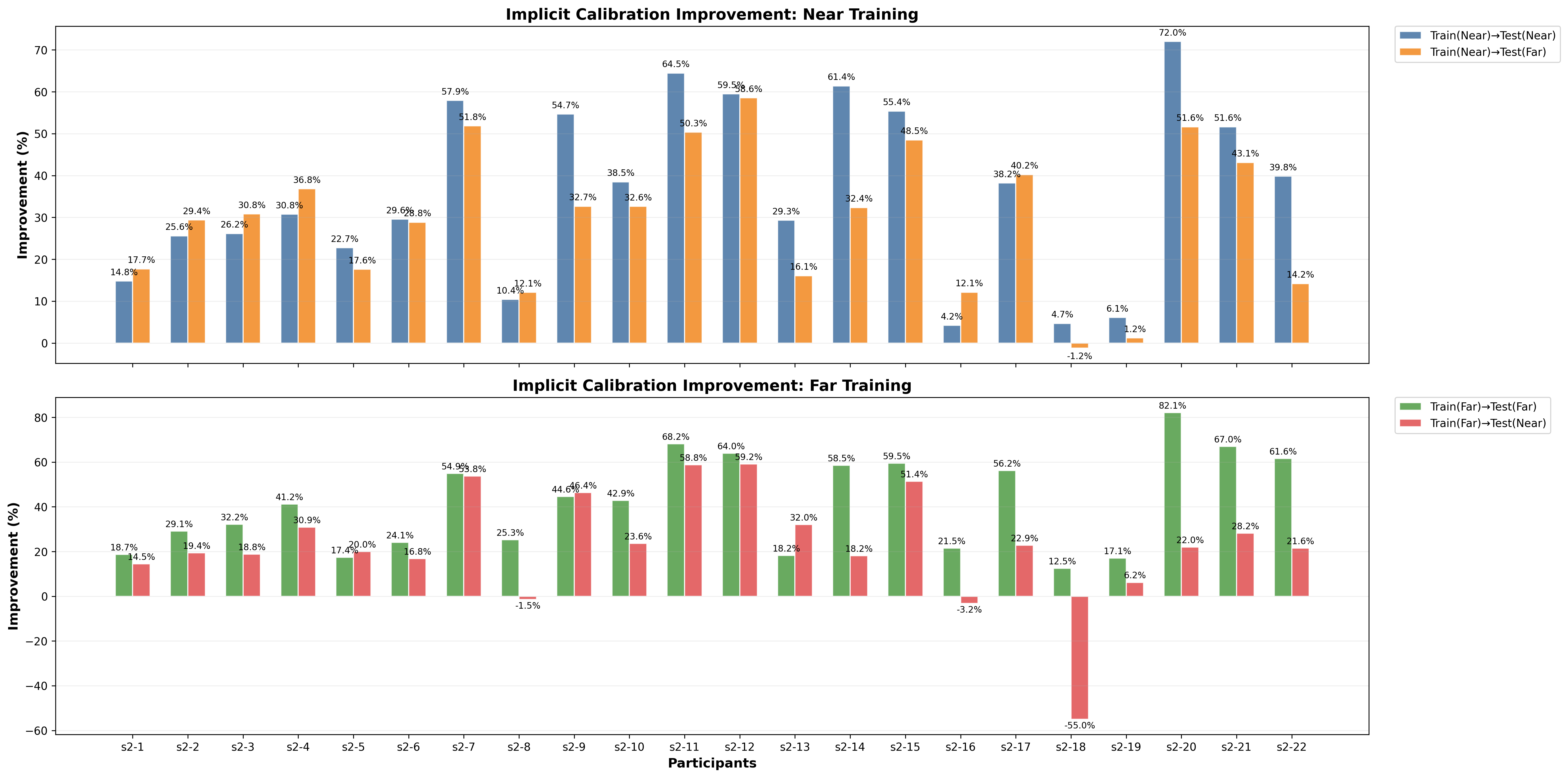}
    \caption{Percentage improvement in gaze error reduction when training models with data from different standing positions. Top: Models trained on data collected with participants standing on the near side as the monitor, tested on gaze data from both near-side and far-side positions. Bottom: Models trained on data collected with participants standing on the far side of the monitor, tested on gaze data from both the far side and near positions.}
    \label{fig:s2standing}
\end{figure*}

Fig.~\ref{fig:s2standing} shows per-participant improvement across the four train-test standing position combinations.
Participant s2-18 shows negative improvement when the model is trained on data from the opposite standing position (i.e., trained on far~$\rightarrow$~tested on near, or trained on near~$\rightarrow$~tested on far). This occurs because s2-18's raw gaze error is already very low ($<100$ px), leaving little room for further reduction. Consequently, any model trained with mismatched or suboptimal data can introduce additional bias, leading to worse performance.

\subsection{Sampling Strategy Comparison}
\begin{figure*}[htbp]
    \centering
    \includegraphics[width=\linewidth]{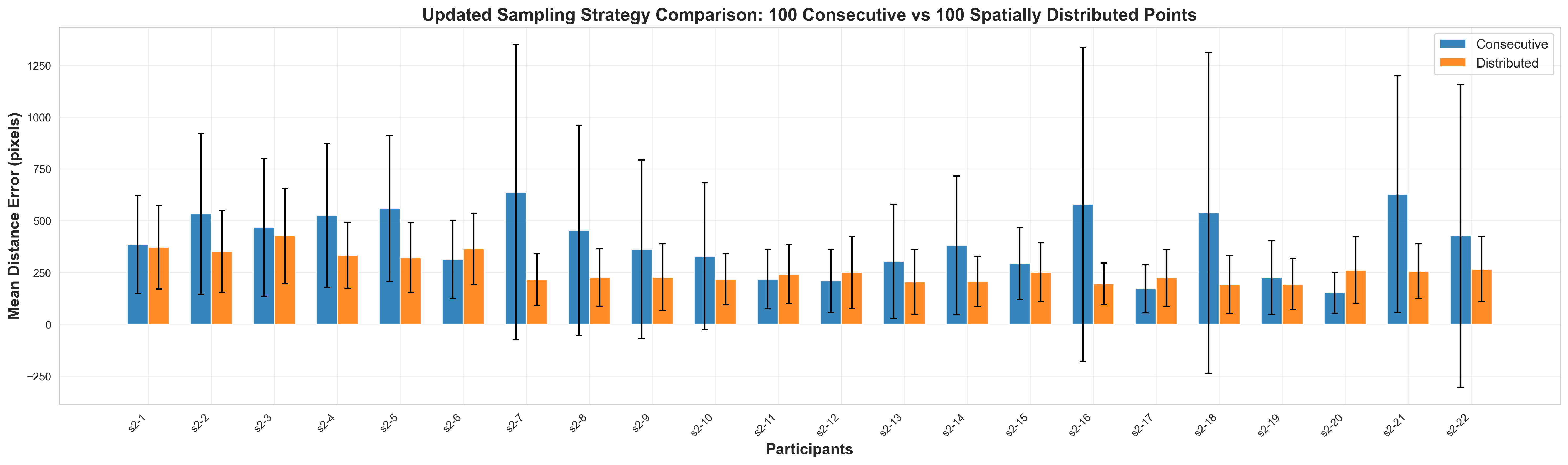}
    \caption{Comparison of sampling strategies for implicit calibration using consecutive versus distributed methods. Distributed sampling yielded consistently lower gaze error across participants, whereas consecutive sampling produced more variable and often higher errors.
}
    \label{fig:s2spatial}
\end{figure*}

Fig.~\ref{fig:s2spatial} compares consecutive versus distributed sampling strategies per participant.
Some of the error bars extend below zero because the plot visualizes mean $\pm$ standard deviation (or SEM). For several participants, the standard deviation is larger than the mean, causing the lower bound of the error bar to fall into negative values, even though the actual distance errors are never negative.

This effect is especially pronounced for participants such as s2-7, s2-16, and s2-21, where the model trained on consecutive-frame samples performed poorly. For these participants, the calibration model occasionally produced extremely large errors, dramatically increasing the variance. As a result, the error distribution becomes highly spread out, and the error bars appear to dip below zero due to the statistical visualization rather than negative distance errors.

\subsection{Sample Size Effect on Calibration Performance}
\begin{figure}[htbp]
    \centering
    \includegraphics[width=0.8\linewidth]{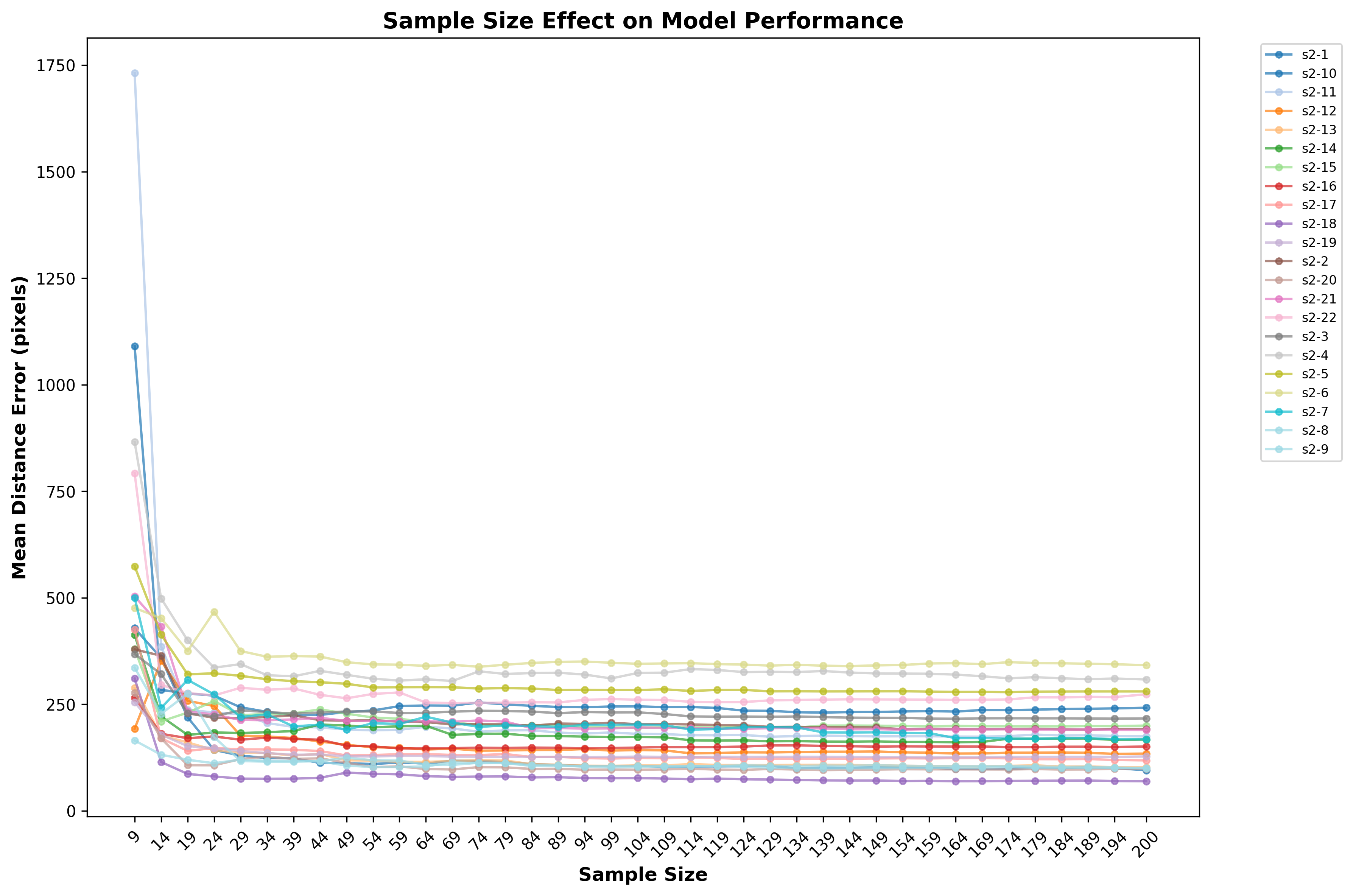}
    \caption{Sample size effect on implicit calibration model performance. Across participants, the mean distance error decreased rapidly with the first few dozen samples, then gradually plateaued.}
    \label{fig:s2samplesize}
\end{figure}

Fig.~\ref{fig:s2samplesize} shows how calibration error decreases as training sample size increases.

\subsection{Regression Model Comparison}
\begin{figure}[htbp]
    \centering
    \includegraphics[width=0.8\linewidth]{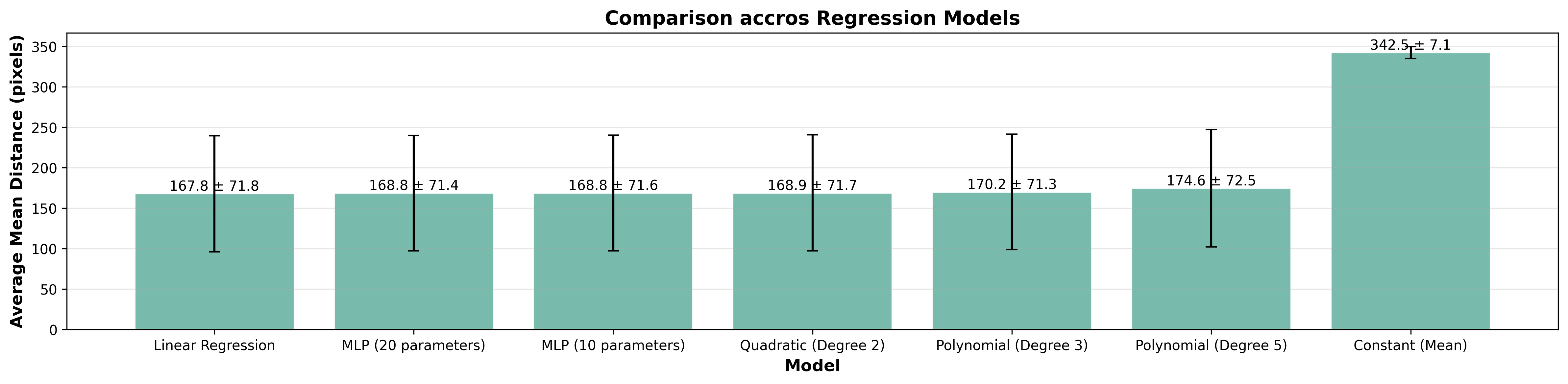}
    \caption{Average performance of different regression models for implicit calibration in Study 2. Linear regression achieves the lowest average error; more complex models (MLPs, polynomial regressions) do not improve results.}
    \label{fig:s2models}
\end{figure}

Fig.~\ref{fig:s2models} compares the performance of different regression models.

\subsection{Convergence Sample Sizes}
Table~\ref{tab:s2convergence} lists the convergence sample size for each participant.
\begin{table}[htbp]
\small
\begin{tabular}{rcc}
\hline
ID & \makecell{Convergence\\Sample Size} & \makecell{Improvements Over\\Last 3 Windows} \\
\hline
s2-20          & 34                      & 0.10\%, -13.53\%, -4.96\%        \\
s2-21          & 44                      & 0.40\%, -0.90\%, -1.81\%         \\
s2-18          & 44                      & 0.36\%, -0.36\%, -2.60\%         \\
s2-14          & 44                      & -1.07\%, -1.51\%, -8.17\%        \\
s2-15          & 44                      & -1.15\%, -0.63\%, -4.23\%        \\
s2-9           & 49                      & 0.04\%, -1.30\%, -6.49\%         \\
s2-3           & 54                      & -1.04\%, -1.05\%, -0.01\%        \\
s2-1           & 54                      & -0.61\%, -2.94\%, -1.59\%        \\
s2-17          & 64                      & -0.90\%, -1.43\%, -0.20\%        \\
s2-19          & 64                      & 0.21\%, -1.18\%, 0.27\%          \\
s2-8           & 69                      & -0.01\%, -1.32\%, -9.16\%        \\
s2-5           & 69                      & -0.12\%, -0.06\%, 0.05\%         \\
s2-13          & 74                      & 0.28\%, -2.11\%, -1.33\%         \\
s2-16          & 74                      & 0.07\%, -0.59\%, -0.60\%         \\
s2-22          & 79                      & 0.48\%, -0.25\%, -0.12\%         \\
s2-12          & 89                      & -0.30\%, -0.78\%, -0.04\%        \\
s2-6           & 89                      & -1.43\%, -1.20\%, -0.74\%        \\
s2-2           & 99                      & -2.57\%, 0.37\%, -1.25\%         \\
s2-7           & 104                     & -2.37\%, -0.03\%, -0.28\%        \\
s2-4           & 114                     & -4.18\%, -0.17\%, -2.60\%        \\
s2-10          & 124                     & -0.80\%, -0.27\%, -0.37\%        \\
s2-11          & 144                     & -0.94\%, -0.50\%, 0.42\%          \\
\hline
\end{tabular}
\caption{Convergence sample sizes for each participant, determined when model error improvements dropped below 1\% across three consecutive sliding windows (window size = 5). The listed percentages show the relative error change in the last three windows before convergence.}
\label{tab:s2convergence}
\end{table}

\section{Appendix for Study 3}
\subsection{Participant Demographic}
Table~\ref{tab:my-table} summarizes the demographics of the five surgeon participants in Study~3.
\begin{table}[H]
\begin{tabular}{rcccc}
\hline
Participant ID & Training Level & Case ID  & Gender & Race  \\
\hline
s3-1           & Resident      & Case \#1 & Female & White \\
s3-2           & Resident       & Case \#2 & Male   & White \\
s3-3           & Attending      & Case \#2 & Male   & White \\
s3-4           & Attending      & Case \#3 & Male   & White \\
s3-5           & Attending      & Case \#4 & Male   & White  \\
\hline
\end{tabular}
\caption{Participants' demographics of Study 3. }
\label{tab:my-table}
\end{table}
\subsection{\textit{SurgGaze} Codebook}
\label{sec:codebook}
We provide this codebook of implementation \textit{SurgGaze} to benefit future researchers.
\textit{Phase 1: Identifying the Calibration Sample}

A critical step of using \textit{SurgGaze} is to identify the calibration sample, i.e., frames in which the surgeon's gaze point on the screen is known with high certainty. 
In the current implementation, we took a guided-manual approach where a medical expert followed a routine to select and annotate frames. In this section, we detail the steps the medical expert took. We hope to present this codebook as a guide for other researchers and practitioners. We will also discuss how this process may be automated in future work. 
Calibration samples are identified based on tool-tissue contact, which provides the most reliable implicit gaze targets.
Study 2 suggests that approximately 150 annotated frames are sufficient for calibration. In Study 3, a medical expert routinely identified and labeled about 150 frames within 15 minutes for each surgery. While our current protocol is manual, it is amenable to automation to improve scalability and to ensure broader spatial coverage across the display. Even as-is, annotating 
\textasciitilde150 frames (\textasciitilde15 minutes) is a modest effort relative to the payoff: a calibration that can support a two-hour procedure (and potentially additional cases by the same surgeon). Here are the specific steps in annotation: 
\begin{enumerate}
    \item Identify the dissection period within the surgery (typically $\sim$20 minutes).
    \item Within this $\sim$20 minutes segment, the annotator aims to select \textasciitilde150 frames where the \textbf{surgical tooltip is in contact with tissue}, indicating the surgeon has a higher likelihood of fixation on the contact point between tooltip and tissue. 
    \begin{itemize}
        \item For ever frame, make a binary decision: \textit{contact} or \textit{no contact}.
        \item If \textit{contact}, check whether it is a initial contact or a burning moment with energy applied on tooltip.
        \item If \textit{yes}, this frame is selected, then designate the tool-tissue contact point as the ground truth gaze location.
    \end{itemize}
\end{enumerate}

In addition to the steps, here are some strategies to identify frames more efficiently:
\begin{itemize}
    \item Single-ended tools, such as the hook or suction device, provide more accurate targets and are recommended. Two-tip tools (e.g., dissector) should be avoided, which introduce ambiguity regarding which end is the active cursor.
    \item During initial tool-tissue contact, the tool (cursor) and surgical target overlap spatially. This provides the most accurate implicit ground truth for gaze.
    \item Energy application moments, e.g., when the hook tip is burning the tissue, are particularly reliable, since careful attention is required to apply energy precisely. The surgical tooltip at the moment of energy delivery can be assumed to represent the surgeon’s gaze with high confidence.
\end{itemize}

In Study 3, we provided a labeling interface\footnote{\url{https://github.com/CVHub520/X-AnyLabeling}} for the medical expert to more quickly annotate the frames.

\textit{Phase 2: Identifying the Testing Sample.}
When applying \textit{SurgGaze} in practice (after seeing success), it does not require the identification of testing frames. For research and experimentation purposes, we also introduce the methods we employed to identify testing frames. Although we could possibly use similar annotation methods as in the training phase to identify testing samples. To approximate real usage, we didn't use the same annotation method to avoid the training and testing samples being too similar. Instead, during the surgery operation, we asked our participating surgeons to verbalize where they were looking. We then reinstalled the specific frames where the surgeons were describing where they were looking. We then defined regions of interest (ROIs) based on surgeons' verbal references. 
This gives us a more diverse testing sample involving a variety of scene compositions and areas of interest. Here are the steps we took:

\begin{enumerate}
    \item We asked the participating surgeons to \textbf{verbalize} or \textbf{point to the monitor} to indicate their gaze location.
    \item During data analysis, we extracted the frames corresponding to these explicit indications (verbal references and explicit on-screen pointing).
    \item We annotate an ROI on the screen as the "Gaze Area", representing where the surgeon reported looking.
\end{enumerate}

Here are some examples of verbal references 
\begin{itemize}
    \item \textbf{Anatomical Landmarks}  
    \begin{itemize}
        \item “Go a little higher, that’s the xiphoid.”
        \item “You’re right on top of the gallbladder here.”  
        \item “That’s just fat, not a vessel.”  
        \item “You’re in the peritoneum, keep opening that up.”  
    \end{itemize}
    \item \textbf{Texture / State}  
        \begin{itemize}
          \item “That’s a nice wispy layer, just peel it away.”  
          \item “This part is really thick; thin it out before you grab.”  
          \item “The tissue looks inflamed, so take your time here.”  
          \item “Now it’s clear, we can see the edge of the duct.”  
          \item “Clean off all that schmutz so the view is better.”  
        \end{itemize}  
   \item \textbf{Planes / Layers}  
    \begin{itemize}
      \item “Work along this lip right here.”  
      \item “You just made a hole, open that up a little more.” 
      \item “Drop down into the fossa and sweep laterally.”  
    \end{itemize}   
\end{itemize}

\subsection{Failed cases}
\begin{figure}[htbp]
    \centering
    \includegraphics[width=\columnwidth]{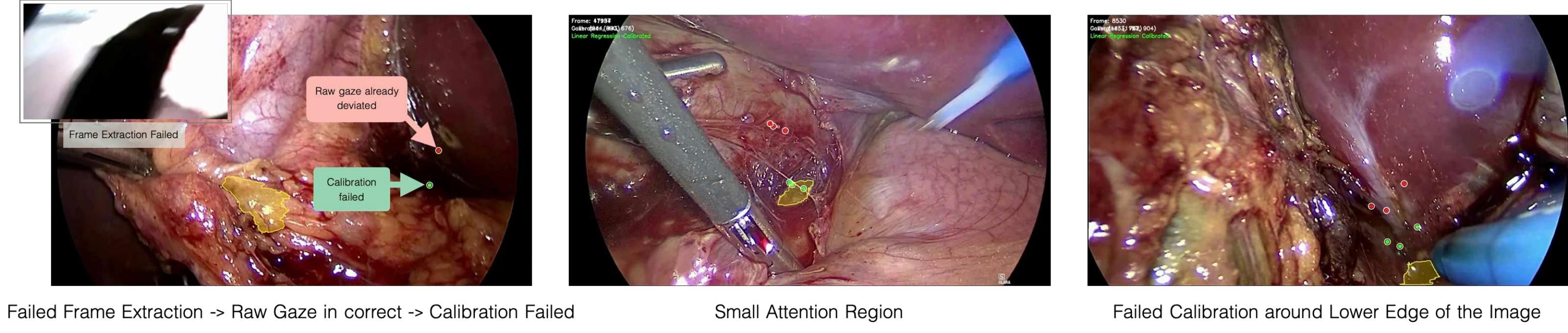}
    \caption{Three failed cases of \textit{SurgGaze} in test data (Study~3). Left: Frame extraction failure. Middle: Very small attention region. Right: Limited training data near screen edges.}
    \label{fig:s3failed}
\end{figure}

We also observed several failure cases where the calibrated gaze does not fall into the region of interest, as shown in Fig.~\ref{fig:s3failed}.

\textbf{Frame recognition errors.} In some cases, raw gaze was not correctly exported, which propagated errors into the calibration step and resulted in misaligned gaze estimates.
\textbf{Very small targets.} When the referenced anatomical structure was too small, even minor gaze deviations caused the calibration to fail in capturing it reliably.
\textbf{Sparse sampling near edges.} For regions at the periphery of the laparoscopic screen, too few calibration points were available around the edges. Although the calibrated gaze was closer to the ground truth than the raw gaze, accuracy remained suboptimal.

\end{document}